\documentclass{elsart}
 
\newcommand{\beq}{\begin{equation}}
\newcommand{\eeq}{\end{equation}}
\newcommand{\beqa}{\begin{eqnarray}}
\newcommand{\eeqa}{\end{eqnarray}}

\newcommand{\ds}{{\sffamily DarkSUSY}}

\usepackage{graphicx}
\usepackage{epsfig} 

\begin{document}

\begin{frontmatter}

\title{The Galactic Center as a Dark Matter Gamma-Ray Source}

\author{Alessandro Cesarini$^1$, Francesco Fucito$^1$, Andrea Lionetto$^1$,} 
\author{Aldo Morselli$^1$ and Piero Ullio$^2$}

\address{$^1$ INFN Roma2  and  University of Roma "Tor Vergata", Via della 
Ricerca Scientifica 00133 Rome, Italy  \\
$^2$ SISSA, Via Beirut 2-4, I-34013 Trieste, Italy, 
and INFN, Sez. di Trieste, Italy}

\maketitle


\begin{abstract} 

The EGRET telescope has identified a gamma-ray source at the Galactic center. 
We point out here that the spectral features of this source are compatible with the 
gamma-ray flux induced by pair annihilations of dark matter weakly interacting 
massive particles (WIMPs). We show that the discrimination between this 
interpretation and other viable explanations will be possible with GLAST,
the next major gamma-ray telescope in space, on the basis of both the spectral
and the angular signature of the WIMP-induced component. If, on the other hand, 
the data will point to an alternative explanation, we prove that there will still
be the possibility for GLAST to single out a weaker dark matter source at the
Galactic center. The potential of GLAST has been explored both in the context
of a generic simplified toy-model for WIMP dark matter, and in a more specific
setup, the case of dark matter neutralinos in the minimal supergravity framework. 
In the latter, we find that even in the case of moderate dark matter densities in the
Galactic center region, there are portions of the parameter space which
will be probed by GLAST. 

\end{abstract}
\end{frontmatter}

\section{Introduction}

Experimental cosmology has been steadily progressing over the latest years.
The emerging picture has been recently reinforced by the data from the Wilkinson 
Microwave Anisotropy Probe (WMAP) which have pinned down several fundamental 
parameters to a remarkable level of precision. In particular, in the latest global
fit\cite{wmap}, the contribution to the critical density of non-relativistic 
matter has been found in the range $\Omega_{m} h^2 = 0.135^{+ 0.009}_{-0.008}$
(here $h$ is the Hubble constant in units of 100 km s$^{-1}$ Mpc$^{-1}$;
$h = 0.71^{+0.04}_{-0.03}$\cite{wmap}), much larger than the baryonic term,
$\Omega_{b}h^2 = 0.0224 \pm 0.0009$.

Unveiling the nature of non-baryonic cold dark matter (CDM) is one of the major 
challenges in science today. Weakly interacting massive particles (WIMPs) are 
among the leading dark matter candidates. They would naturally appear as another 
of the thermal leftovers from the early Universe, and, at the same time, their 
existence is predicted in several classes of extensions of the Standard Model of 
particle physics. The most popular of such candidates is the lightest
neutralino in R-parity conserving supersymmetric models. Considerable effort has 
been put in the search for dark matter WIMPs in the last decade, with several 
complementary techniques applied (for a recent review, see, e.g.,~\cite{lars}).
Among them, indirect detection through the identification of the yields
of WIMP pair annihilations in dark matter structures\cite{silksrednicki,stecker}
seems to be a very promising method. In particular, we will focus here, as a 
signature to identify dark matter, on the possible distortion of the spectrum of 
the diffuse $\gamma$-ray flux in the Galaxy due to a WIMP-induced component, 
extending up to an energy equal to the WIMP mass (a list of other recent analysis 
on this topic includes Refs.~\cite{p1,p2,p3,p4,p5,p6,p7}).

The EGRET telescope on board of the Compton Gamma-Ray Observatory
has mapped the $\gamma$-ray sky up to an energy of about 20~GeV
over a period of 5 years. The data collected by EGRET toward the Galactic center (GC) 
region show~\cite{Mayer} high statistical evidence for a gamma-ray 
source, possibly diffuse rather than point-like, located within $1.5^\circ$ of the GC 
($l=b=0^\circ$). The detected flux largely exceeds the diffuse $\gamma$-ray 
component expected in the GC direction with a standard modeling of the interaction 
of primary cosmic rays with the interstellar medium  (see, e.g., \cite{smapj}); the
latter fails also to reproduce the spectral shape of the GC source. Although other 
plausible explanations have been formulated, it is very intriguing 
that the EGRET GeV excess shows, as basic features, the kind of distortion of the 
diffuse $\gamma$-ray spectrum one would expect from a WIMP-induced component, assuming
that the dark matter halo profile is peaked toward the GC. We will identify for 
which classes of WIMP compositions and masses, fair agreement with the measured flux 
can be obtained.

No firm conclusion about the nature of the GC excess can be driven from data available 
at present; on the other hand, the picture is going to become much clearer in the near 
future. The Gamma-ray Large Area Space Telescope (GLAST)\cite{glast}
has been selected by NASA  as the next major $\gamma$-ray mission,\footnote{A list of 
people and institutions involved in the collaboration together with the on-line status 
of the project is available at {\sl http://www-glast.stanford.edu}.
For a detailed description of the apparatus see~\cite{Bellazzini};
a discussion of the main scientific goals can be found in~\cite{morselli}.} 
and is scheduled for launch in the first half of 2006. Compared to EGRET, GLAST will have 
a much larger effective area, better energy and angular resolutions, as well as it will 
cover a much wider energy range. GLAST will perform an all-sky survey of $\gamma$-ray 
sources, with scientific objectives including the study of
blazars, $\gamma$-ray bursts, supernova remnants, pulsars, the diffuse 
radiation in the Galaxy, and unidentified high-energy sources.
The identification of dark matter sources has been indicated 
as one of its main scientific goals. We illustrate here the 
conditions under which it may be feasible that GLAST will single out
a dark matter source located at the GC.

The paper is organized as follows. In Section~\ref{foed} we show that the GC EGRET 
$\gamma$-ray excess can be modeled with a component due to WIMP annihilations. 
In Section~\ref{egretglast} we discuss what kind of information GLAST could provide
to confirm such hypothesis. In Section~\ref{glastweak} we illustrate the perspectives of 
dark matter detection with GLAST in the case in which the EGRET excess will be found 
to be due to another kind of source. In the first part of the paper
we discuss features for the dark matter signal in the contest of a 
generic toy-model; in Section~\ref{sugra} we will focus on neutralino dark matter
candidates in the minimal supergravity framework, applying to this specific case
the tools developed in the other Sections. Conclusions are in Section~\ref{concl}.

\section{A Dark Matter Source at the Galactic Center?}
\label{foed}

\begin{table}[t]
\begin{center}
\caption{Estimated values for the Galactic diffuse $\gamma-$ray 
component (second column) and EGRET data from a region of $1.5^\circ$ around 
the GC (third column), extracted from \cite{Mayer}.}
\vspace{0.5cm}
\begin{tabular}{ccc}
\hline
Energy Bin & Expected Diffuse $\gamma-$Ray Flux  &Total $\gamma-$Ray Flux  \\
$\left(\rm{GeV}\right)$ & 
$\left(\rm{cm}^{-2}\rm{s}^{-1}\rm{GeV}^{-1}\rm{sr}^{-1}\right)$ &
$\left(\rm{cm}^{-2}\rm{s}^{-1}\rm{GeV}^{-1}\rm{sr}^{-1}\right)$ \\
\hline
 $ 0.03-0.05$ & $3.7\cdot 10^{-3}$ & $(5.0\pm0.8)\cdot 10^{-2}$ \\
 $ 0.05-0.07$ &   $1.8\cdot 10^{-3}$ & $(1.3\pm0.2)\cdot 10^{-2}$ \\
 $ 0.07-0.1$ & $1.1\cdot 10^{-3}$ & $(6.1\pm0.5)\cdot 10^{-3}$ \\

 $ 0.1-0.15$ & $6.2\cdot 10^{-4}$ & $(4.4\pm0.2)\cdot 10^{-3}$ \\

 $ 0.15-0.3$ & $2.6\cdot 10^{-4}$ & $(2.03\pm0.06)\cdot 10^{-3}$ \\

 $ 0.3-0.5$  & $1.0\cdot 10^{-4}$ & $(9.5\pm0.2)\cdot 10^{-4}$ \\

 $ 0.5-1$  & $3.5\cdot 10^{-5}$ & $(3.9\pm0.1)\cdot 10^{-4}$ \\

 $ 1-2$ & $9.1\cdot 10^{-6}$ & $(1.52\pm0.03)\cdot 10^{-4}$ \\
 $ 2-4$  & $2.0\cdot 10^{-6}$ & $(3.2\pm0.1)\cdot 10^{-5}$ \\
  $4-10$ & $2.3\cdot 10^{-7}$ & $(3.1\pm0.2)\cdot 10^{-6}$ \\
\hline
\end{tabular}
\label{egretdata}
\end{center}
\vspace{1.cm}
\end{table}

In Table~\ref{egretdata} we report the flux per energy bin for the GC gamma-ray 
source as measured by EGRET, together with the expected flux from cosmic ray 
interactions in a standard scenario~\cite{Mayer,Mayerpriv} 
(see also Table~2 and Fig.~4 in~\cite{Mayer}). As already mentioned there is 
a significant mismatch between the two: we entertain here the possibility
that the bulk of the high energy flux is due to pair annihilations of dark matter
WIMP's in the GC region. Hence, we assume that the total flux measured by EGRET can be 
described as the superposition of two contributions:
\begin{itemize}
\item i) a component due the interaction of primary cosmic rays with the
 interstellar medium, with spectral shape defined by the function 
${ S}_b(E_\gamma)$ (background contribution)
\item ii) a component due to WIMP annihilation in the dark matter halo,
whose energy spectrum is defined by ${ S}_\chi(E_\gamma)$ (signal contribution).
\end{itemize}
We write the total $\gamma$-ray flux as:
\begin{equation}
\phi_\gamma= \phi_b + \phi_\chi = N_b { S}_b + N_\chi { S}_\chi,
\label{totflux}
\end{equation} 
where $N_b$ and $N_\chi$ are dimensionless normalization parameters, respectively, 
for the standard and exotic flux, which we define below.

\subsection{The background component}

There are three mechanisms which give rise to diffuse $\gamma$-ray
radiation in the Galaxy: production and decay of $\pi^{0}$s, inverse Compton 
scattering and bremsstrahlung~(see, e.g., \cite{smapj}). According to
standard scenarios, in the energy range we will consider, $E_\gamma>1\;{\rm Gev}$,
the dominant background source is given by $\pi^{0}$ decays. Photons are generated 
in the interaction of primary cosmic rays
with the interstellar medium via:
\[
p+X\to \dots\to\pi^{0}\to2\gamma  
\]
\[
He+X\to \dots \to\pi^{0}\to2\gamma\;,
\]
where $X$ stands for an interstellar atom, mainly $H$ and $He$. We have simulated 
the induced $\gamma-$ray yield according to standard treatments (see, 
e.g.,\cite{stecker-pi0,gaisserbook}) and as implemented in the Galprop software 
package~\cite{smapj}. We assume that the $p$ and $He$ cosmic ray fluxes in the 
Galaxy have the same energy spectra and relative normalization as those measured 
in the local neighborhood~\cite{bess,ams}, and that the $He$ fraction of the 
interstellar gas is 7\%. We write the background flux, splitting it into two factors:
\begin{equation}
  { S}_{b}(E_\gamma)=\frac{1}{\left(1\; \rm{cm}^2 \rm{sr}\right)}\cdot 
  { Em}(E_\gamma)
\end{equation}
and
\begin{equation} 
  N_b=\frac{1}{\left(1\; \rm{cm}^{-2}\rm{sr}^{-1}\right)}\cdot
  \int_{l.o.s.} dl \frac{n_H(l)}{4\pi} 
  \frac{\phi^{prim}_{p}(l)}{\phi^{prim}_{p}(l=0)}\;.
\label{nback}
\end{equation} 
Here ${ Em}(E_\gamma)$ [GeV$^{-1}$ s$^{-1}$] is the local emissivity per hydrogen
atom, i.e. the number of secondary photons with energy in the range 
($E_\gamma$, $E_\gamma + dE_\gamma$) emitted per unit time per target hydrogen
atom, for an incident flux of protons and helium nuclei equal to
the locally measured primary proton and helium fluxes. The factor 
$N_{b}$  is instead associated to the interstellar hydrogen column density 
$n_H(l)$, integrated along the line of sight and weighted over the proton primary 
flux at the location $l$, $\phi^{prim}_{p}(l)$, normalized to the local value 
$\phi^{prim}_{p}(l=0)$. 

Above an energy of about 1~GeV the background spectrum (and therefore the function
$\phi_{b}$) can be described by a power law of the form $\phi_{back} \propto E_\gamma^{-\alpha}$ with the same spectral index as the dominant primary component,
i.e. the proton spectral index $\alpha = 2.72$.

The relative normalization of the primary components in different places in the 
Galaxy can be estimated once a radial distribution of primary sources is chosen
(following, for instance, the radial distribution of supernovae) and then
by propagating the injected fluxes with an appropriate transport equation
(this is what is done, e.g., in the Galprop code~\cite{smapj}). On the other 
hand, the hydrogen column density toward the Galactic center is very uncertain;
we choose therefore to define the spectral shape of the background through the 
function ${ S}_b$, and to keep $N_b$ as a free normalization parameter.

\subsection{The signal component}

The production of $\gamma$-rays in a dark matter halos made of WIMP's follows
essentially by the definition of WIMP, regardless of any specific scenario
one has in mind. The signal scales linearly with the pair annihilation rate
in the limit of non-relativistic particles. 

We consider a generic framework in which the dark matter in the Galactic halo
is made of  particles $\chi$, WIMP dark matter candidates with mass 
$M_\chi$ and total pair annihilation rate into lighter 
Standard Model particles $\sigma v$ (in the limit of zero relative velocity). 
Among the kinematically-allowed tree-level final states, the leading channels 
are often $b\bar{b},t\bar{t},\tau^+\tau^-,W^+W^-,Z^0Z^0$. 
This is the case, e.g., for neutralinos and, more generically, for any Majorana 
fermion WIMP, as for such particles the S-wave annihilation rate into the light 
fermion species is suppressed by the factor $m_{f}^2/M_\chi^2$, where $m_{f}$ 
is the mass of the fermion in the final state. The fragmentation and/or the decay 
of the tree-level annihilation states gives rise to photons. Again the dominant 
intermediate step is the generation of neutral pions and their decay into $2\gamma$. 
The simulation of the photon yield is standard; we take advantage of a simulation
performed with the Lund Monte Carlo program Pythia 6.202~\cite{pythia}
implemented in the \ds\ package~\cite{ds}.

Suppose that the dark matter halo is roughly spherical and consider the 
induced $\gamma$-ray flux in the direction that forms an angle $\psi$ with
the direction of the Galactic center; the WIMP induced photon flux is the sum 
of the contributions along the line of sight (l.o.s):
\begin{equation} 
  \phi_\chi(E,\psi)=\frac{\sigma v}{4\pi} \sum_f \frac{dN_f}{dE} B_f
  \int_{l.o.s} dl(\psi) \frac{1}{2}\frac{\rho(l)^2}{M_\chi^2} 
\label{gammafluxcont}
\end{equation}
where $B_f$ is the branching ratio into the tree-level annihilation final state $f$, 
while ${dN_f}/{dE}$ is the relative differential photon yield.
The WIMP mass density along the line of sight,  $\rho(l)$, enters critically
in the prediction for the flux, as the number of WIMP pairs is equal to 
$1/2\,\rho(l)^2/M_\chi^2$. 
It is then useful to factorize the flux in Eq.(\ref{gammafluxcont}) into two pieces, 
one depending only on the the particle physics setup, i.e. on the cross section, 
the branching ratios and the WIMP mass, and the other depending on the WIMP 
distribution in the galactic halo. We rewrite Eq.(\ref{gammafluxcont}) as~\cite{bub}:
\begin{eqnarray}
  \phi_\chi(E,\psi) & = & 3.74 \cdot 10^{-10}\left( \frac{\sigma v}{10^{-26}\;
  \rm{cm}^3 \rm{s}^{-1}}\right)\left( \frac{50\; \rm{GeV}}{M_\chi}\right)^2 
  \sum_f \frac{dN_f}{dE} B_f \nonumber \\
  & & \cdot  J(\psi) \;\rm{cm}^{-2} \rm{s}^{-1} \rm{GeV}^{-1} \rm{sr}^{-1} 
\label{wimpflux}
\end{eqnarray}
where we introduced the dimensionless function $J$, containing the dependence 
on the halo density profile,
\begin{equation} 
  J(\psi)=\frac{1}{8.5\;\rm{kpc}} 
  \left(\frac{1}{0.3 \;\rm{GeV} \rm{cm}^{-3}}\right)\int \rho^2 (l)dl(\psi)
\end{equation}
More precisely, given a detector with angular acceptance $\Delta\Omega$, we 
have to consider the average of $J(\psi)$ over the solid angle $\Delta\Omega$
around the direction $\psi$:
\begin{equation}
  \langle J(\psi) \rangle_{\Delta\Omega}=\frac{1}{\Delta\Omega}\int J(\psi)d\Omega
  \label{jpsiave}
\end{equation}
To compare with the GC EGRET gamma-ray source, we will consider
$\Delta\Omega\sim 10^{-3}$~sr, i.e. the same magnitude as the angular region
probed by the EGRET telescope.

\begin{figure}[t]
\begin{center}
\includegraphics[scale=0.20]{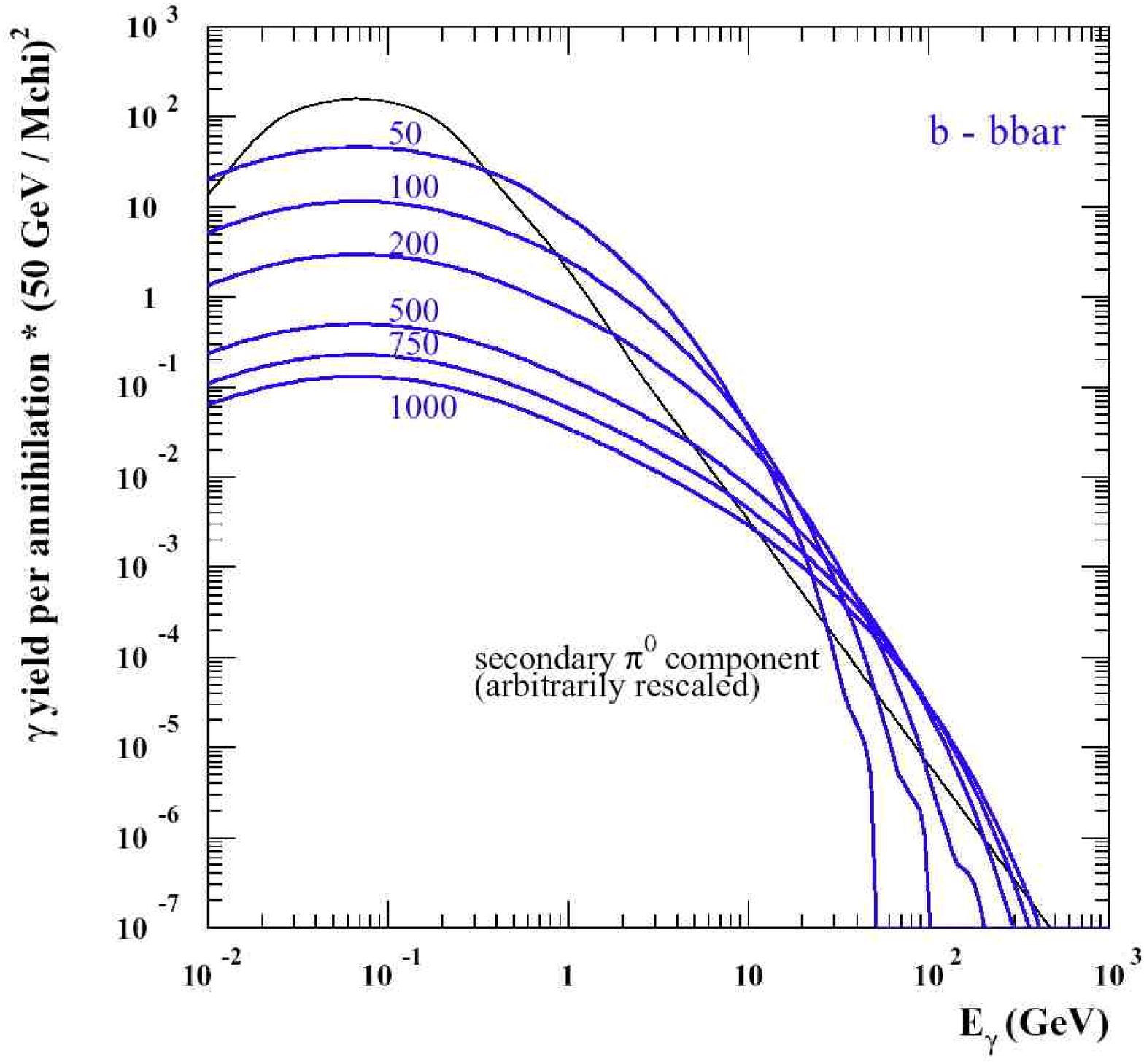}
\includegraphics[scale=0.20]{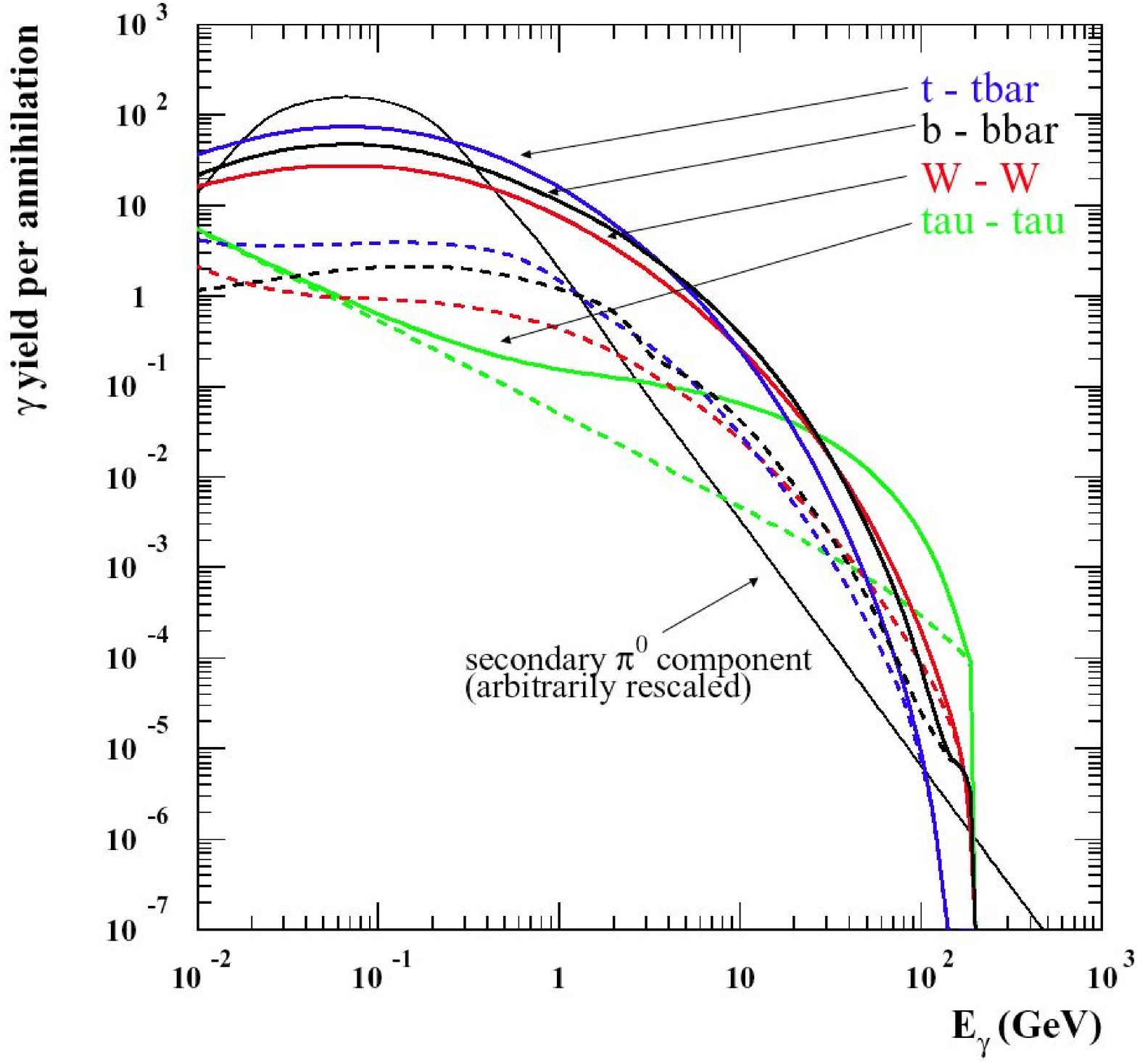}
\caption{In the left panel: differential $\gamma$-ray yield per annihilation 
(see Eq.(\ref{gammafluxcont})) for 
a fixed annihilation channel ($b\bar{b}$) and for a few sample values of WIMP masses. 
For comparison we also show the emissivity, 
with an arbitrarily rescaled normalization, from the interaction of primaries with 
the interstellar medium. In the right panel: differential yield per 
annihilation for a few sample annihilation channels and a fixed WIMP mass 
($200\; \rm{GeV}$). The solid lines are the total yields, while the dashed lines are 
components not due to $\pi^0$ decays.}
\label{fign1}
\end{center}
\end{figure}

As for the background component, we split the signal into a term
which fixes the spectral shape of the flux, plus a normalization factor.
In the notation introduced in Eq.(\ref{totflux}), we label
$N_\chi \equiv \langle J(\psi) \rangle_{\Delta\Omega}$ and define
${ S}_\chi \equiv \phi_\chi / N_\chi$. The dependence on $\rho(l)$ has been
included in the term we treat as a free normalization parameter $N_\chi$,
as $\rho(l)$ is very uncertain both from the theoretical and the observational
points of view. Although there is a some spread in the
predictions for the $\gamma$-ray flux when coming to specific WIMP models, 
its spectral features are rather generic. As most photons are produced in the 
hadronization and decay of $\pi^0$s, the shape of the photon spectrum is always
peaked at half the mass of the pion, about 70 MeV, and it is symmetric around 
it on a logarithmic scale (sometimes this feature is called the ``$\pi^0$ bump'', 
see, e.g.,~\cite{stecker-pi0}). The same is true for the background, but still
it may be possible to tell signal from background: the signal arises in
processes which have all the same energy scale, i.e. $2 M_\chi$, therefore the 
WIMP induced flux, contrary to the background, is spectral index free and shows a sharp
cutoff when $E_\gamma$ approaches the WIMP mass. This is shown in the left panel of
Fig.~\ref{fign1}, where we plot the differential photon yield per annihilation
times the inverse of WIMP mass squared, for a few values of the WIMP mass, and 
assuming that the $\chi$'s have a single dominant annihilation channel ($b\bar{b}$ in the case 
displayed). In the same figure, for comparison, the spectral shape of the 
background is shown: as it can be seen, one may hope to identify the
WIMP induced component as a distortion of the background spectrum at relatively
high energies.
For a given WIMP mass, the photon yields in the different annihilation channels 
are analogous, as shown in the right panel of Fig.~\ref{fign1}: solid curves
indicate the total photon yield, while dashed curves indicate the photon yield
in radiative processes, i.e. in all processes rather than $\pi^0$ decays. 
The spectrum for the $t\bar{t}$ and $W^+W^-$ channels are very close to the one 
for $b\bar{b}$ (differences are mainly given by prompt decays before hadronization);
only in the $\tau^+\tau^-$ case, radiative photon emission is dominant, still
with a large bump due to the hadronic decay modes of $\tau$ leptons.

\subsection{EGRET data fit}
\label{fitsub}

\begin{figure}[t]
\begin{center}
\includegraphics[scale=0.28]{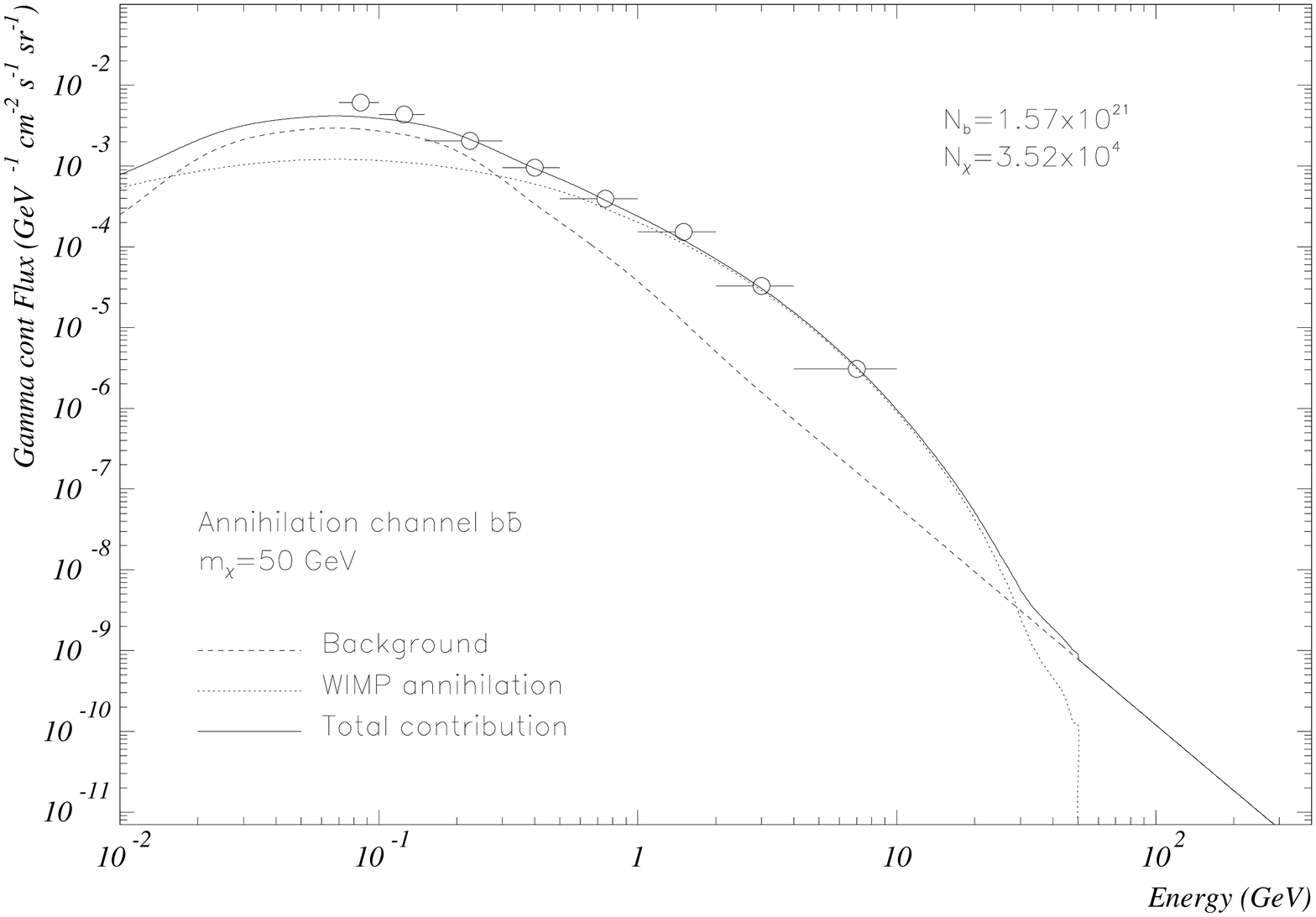}
\includegraphics[scale=0.28]{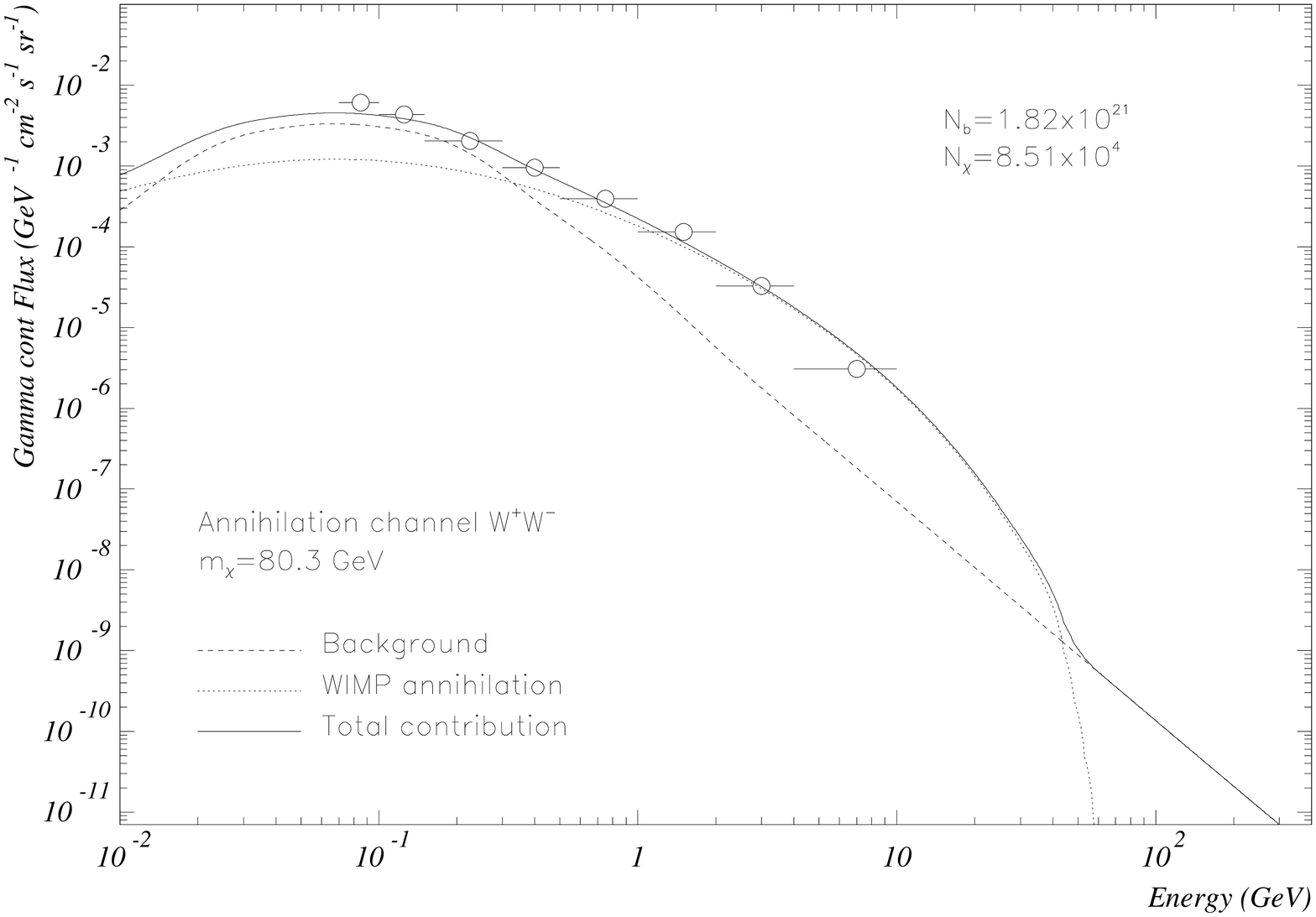}
\end{center}
\caption{Fit of the EGRET GC $\gamma$-ray data for two sample WIMP models.
We fix the WIMP mass ($M_\chi = 50$~GeV in the upper panel, $M_\chi = 80.3$~GeV
in the lower panel) and select a single annihilation channel in each of the
two cases ($b\bar{b}$ in the upper panel, $W^-W^+$ in the lower one). Signal and 
background components are indicated separately, while their sum is shown with a solid 
line. For both models the value of the reduced statistical $\chi^2$ variable obtained from the fit
is around 5.}
\label{egretfit}
\end{figure}

EGRET has performed measures in the energy range $30\;{\rm MeV}\div 10\;{\rm GeV}$ with few
bins in the high energy region.
Given the paucity of the data in the highest end of the energy region
in a first approximation it is not sensible
to discriminate different annihilation channels leading to photons in the final state.

It is then convenient to keep the
discussion as general as possible and consider a simplified context 
(a \emph{toy-model}) with a thermal relic $\chi$  for which
only one intermediate annihilation channel is open ($B_f=1$ in that channel).
Furthermore we assume the total annihilation cross section
to scale with the inverse of the relic abundance 
$\Omega_{\chi}$~\cite{lee-weinberg,jkg}:
\begin{equation}
  \sigma v \sim \langle \sigma v \rangle \sim 
  \frac{3\cdot 10^{-27} \rm{cm}^3 \rm{s}^{-1}}{\Omega_{\chi} h^2} \sim 
  3\cdot 10^{-26} \rm{cm}^3 \rm{s}^{-1}\;,
\label{eq:scal}
\end{equation}
where  $\langle \sigma v \rangle$ is the thermally averaged annihilation cross section.

Eq.(\ref{eq:scal}) is not valid in presence of resonances or thresholds near the kinematically 
released energy in 
the annihilation $2M_\chi$ and of coannihilation effects.
In the presence of these conditions we can have large deviations from this
approximate scaling.

Moreover there are cases in which the inverse proportionality between $\Omega_\chi$
and the annihilation rate only gives a lower bound for the latter since a 
non-thermal relic component can provide the relic density needed 
to account for $\Omega_{CDM}$\cite{anomsu}.

In our toy-model the WIMP mass is kept as a free parameter , as we have 
shown that the photon spectrum is rather sensitive to it. The results we discuss below 
are then dependent on a mass scale and an overall normalization parameter and can be 
easily rescaled to fit an explicit model for which $M_\chi$ and $\sigma v$ are specified.

\begin{figure}[t]
\begin{center}
\includegraphics[scale=0.36]{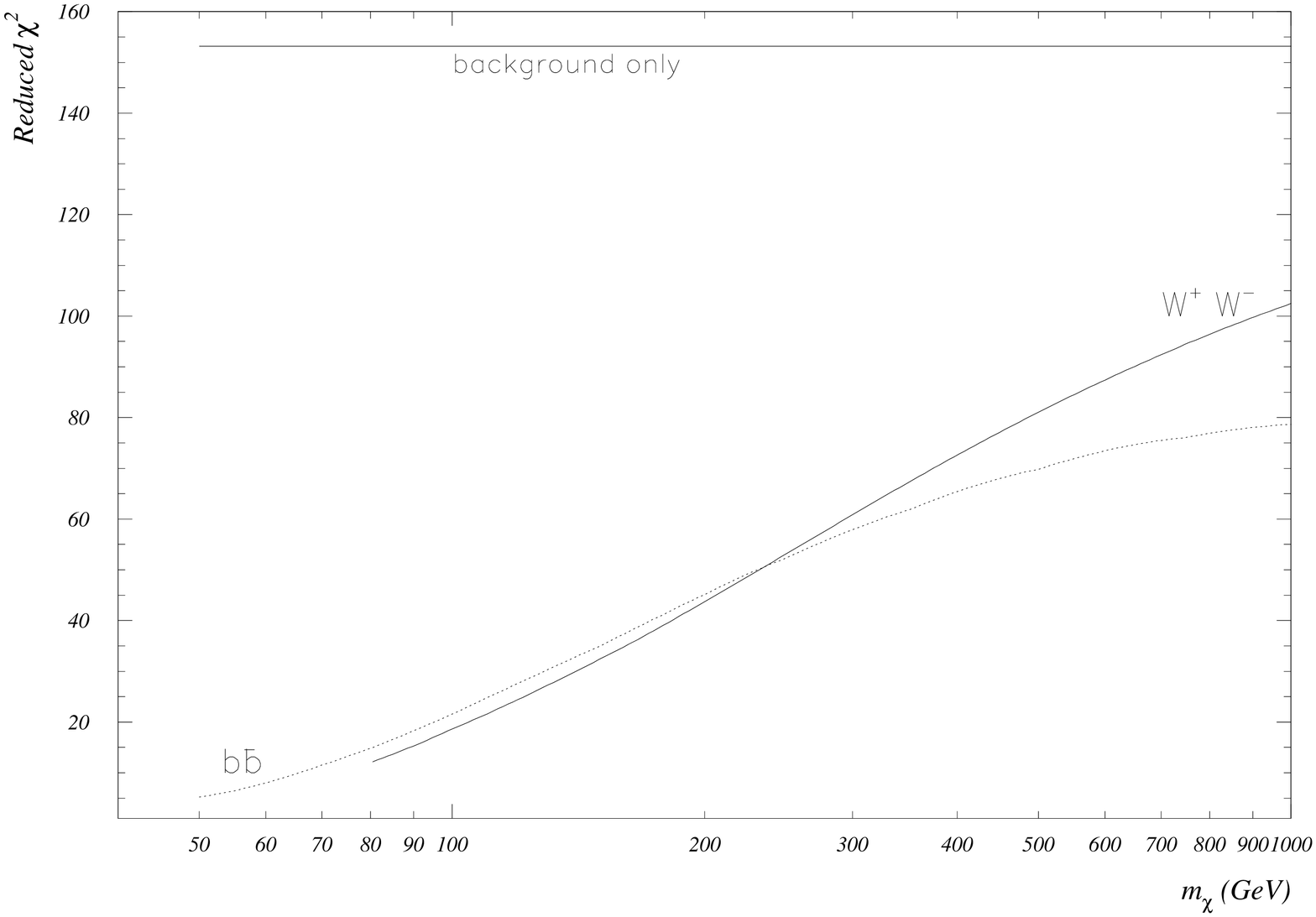}
\caption{Reduced $\chi^2$ corresponding to the best fits of the EGRET excess
for a WIMP model with fixed mass $M_\chi$ and a single annihilation channel 
allowed ($b\bar{b}$ is shown with a dotted line, $W^-W^+$ with a solid line);
the $\chi^2$ has been minimized over the normalizations of both the signal and
the background (with restrictions as explained in the text). Reduced $\chi^2$
values should be compared with the value obtained in case data are fitted
with a background component only, marked with a horizontal line in the upper 
part of the figure.}
\label{fign2}
\end{center}
\end{figure}

We then fit the EGRET data on the GC excess taken from Table~\ref{egretdata} using our
toy-model.
We are not using the two lowest energy bins as for $E\ll 1$~GeV
the background is most probably dominated by inverse Compton and bremsstrahlung
rather than by $\pi^0$ production as we assumed.
We repeat this fit for different values of the WIMP mass 
$M_\chi$ and for a few tree-level annihilation channels. $N_b$ and $N_\chi$ in
Eq.(\ref{totflux}) are treated as free parameters which are varied in order to minimize 
the statistical $\chi^2$ variable. 
The allowed range of variation for the background
normalization, $N_{b}$, is between $3.2\cdot 10^{20}$ and $1.8\cdot 10^{21}$. 
The two extrema of this variation interval are taken to correspond to a best fit of the background in 
agreement 
with the standard scenario (column 2 in Table~\ref{egretdata}) and to the best fit for the EGRET data from
the GC (column 3 in Table~\ref{egretdata}) with $N_\chi=0$. 

In Fig.\ref{egretfit} we show our best fits 
for two sample values of $M_\chi$ and two intermediate channels, 
$b\bar{b}$ and $W^+W^-$. On the qualitative side, the agreement with the data is rather
good, even if the reduced $\chi^2$ (for 6 degrees of freedom) in the two examples 
displayed is still rather large (of the order of 5). This may depend on an underestimate of
error bars or also on the fact that we are neglecting uncertainties in the theoretical
predictions for the spectral shapes. It is clear, on the other hand, that adding
a component due to WIMP annihilations on top of the background component  greatly
improves the agreement between the expected spectral form and the one found in the
EGRET measurement. This is shown in Fig.~\ref{fign2} where the reduced $\chi^2$
for our best fits is shown as a function of WIMP mass.
These values should be compared with those obtained when only the background contribution 
is included ($N_\chi=0$).
This case is represented by the horizontal line in the figure
and gives a reduced $\chi^2\approx 150$. Fig.~\ref{fign2} also indicates that light
WIMP masses are marginally favored over heavier WIMP's. Results for other tree-level 
annihilation states are analogous and show the same trends.

\begin{figure}[t]
\begin{center}
  \includegraphics[scale=0.9]{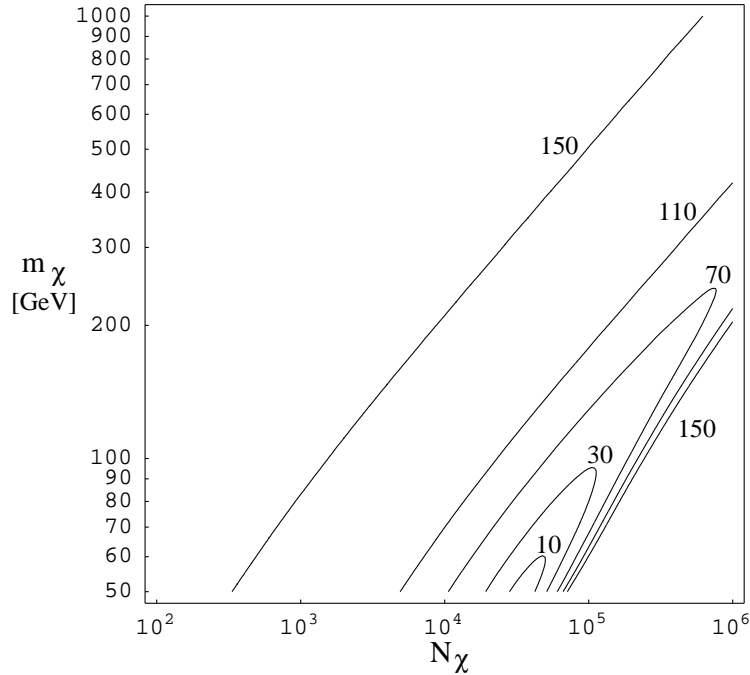}
\end{center}
  \caption{Lines of constant reduced $\chi^2$ corresponding to best fits 
of the EGRET GC excess, in the plane WIMP mass $M_\chi$ versus normalization
of the WIMP-induced signal $N_\chi$. The plot applies to
our toy-model with a single annihilation channel allowed, i.e. $b\bar{b}$ in the
case displayed.}
\label{isochi2}
\end{figure}

The best fit curves displayed correspond to rather large values of the 
normalization parameter $N_{\chi}$, a few times $10^4$ for the two cases shown in 
Fig.~\ref{fign2}. $N_{\chi}$ tends to increase going to heavier WIMP's
as a function inversely proportional to the WIMP mass raised to a certain power.
This power is a little smaller
than the value 2 one would have inferred from the fact that the WIMP pair density decreases 
as $M_\chi^{-2}$.
An investigation of the dependence of the statistical variable $\chi^2$ from the WIMP 
mass and $N_\chi$ is shown in Fig.~\ref{isochi2}, for the $b\bar{b}$ annihilation channel.
Given that $\chi^2=\chi^2(N_b,N_\chi,M_\chi)$, we minimize it  with respect to the parameter $N_b$
for fixed values of $N_\chi$ and $M_\chi$. The isolevel 
curves for the reduced $\chi^2$ are plotted in Fig.~\ref{isochi2}.   
Fixing a value for the WIMP mass and going from the left-hand side to the 
right-hand side of the figure, we move from the case in which the WIMP signal is marginal 
with respect of the background flux to the case where the sum of the two reproduces most 
closely the data. Further increasing the value of $N_\chi$ we reach a region
where the WIMP signal exceeds the flux detected by EGRET.

We recall that in our toy-model  $N_\chi$ is identified with the halo model dependent 
function $\langle J(\psi) \rangle_{\Delta\Omega}$ for the Galactic center direction 
$\psi = 0^{\circ}$ and EGRET angular acceptance $\Delta\Omega = 10^{-3}$~sr.
As already mentioned, the distribution of dark matter in the inner part of the 
Galaxy is still a controversial issue. Dynamical measurements show that
dark matter is, in the Galactic center region, just a subdominant component 
with respect to baryonic matter, but lack the resolution we need for an estimate
of $\langle J \rangle$. 
On the other hand, N-body simulation of structure formation in a CDM Universe,
find dark matter density profiles which are singular toward the center of the Galaxy, 
possibly scaling as $1/r$ (profile of Navarro, Frenk \& White~\cite{nfw} (NFW)) 
or $1/r^{1.5}$
(Moore et al.~profile~\cite{moore2}) as the galactocentric distance 
$r\rightarrow 0$. The corresponding values of $\langle J \rangle$, and hence of 
$N_\chi$ are
listed in the second column of Table~\ref{haloprof} assuming, as commonly done, that the 
dark matter density at the Sun's galactocentric distance is equal to 
$0.3\,\rm{GeV}\,\rm{cm}^{-3}$.
One should however keep in mind that it has been questioned whether
the NFW and the Moore et al. profiles can be used to describe 
inner dark matter halos (especially for smaller galaxies, 
see, e.g., \cite{primack} for a review).
These halo models give a 
snapshot of the Galaxy before the baryon infall. The appearance of a massive black 
hole at the Galactic center and of the stellar components may sensibly modify 
such pictures with further enhancements (but a depletion is possible as well) of the 
central dark matter density~\cite{gs,bhullio}. For comparison, in Table~\ref{haloprof} we 
give the value of $\langle J \rangle$ for the modified isothermal sphere profile, 
which is 
non-singular toward the Galactic center and, as well known, give a normalization for 
the dark matter induced fluxes well below the background and the sensitivity of 
even next-generation detectors. All three halo profiles listed in the Table are consistent
with available dynamical constraints on the Galaxy. We conclude then that $N_\chi$ can be 
at the level needed in our toy-model to reproduce the EGRET excess, but at the same 
time that both larger or smaller values are feasible as well.

\begin{table}
\begin{center}
\caption{Values of $\left< J(0)\right>_{\Delta\Omega}$ for two different 
$\Delta\Omega$'s and for three different density profiles, see the text for details.}
\vspace{0.5cm}
\begin{tabular}{ccc}
\hline 
Profile & $\left< J(0)\right>_{\Delta\Omega}$
($\Delta\Omega=10^{-3}$~sr) & $\left< J(0)\right>_{\Delta\Omega}$
($\Delta\Omega=10^{-5}$~sr)\\
\hline 
Navarro, Frenk, White & $1.21\cdot 10^3$ & $1.26\cdot 10^4$ \\ 
Moore \emph{et al.} & $1.05\cdot 10^5$ & $9.46\cdot 10^6$ \\
Modified isothermal & $3.03\cdot 10^1$ & $3.03\cdot 10^1$ \\        
\hline
\end{tabular}
\vspace{1cm}
\label{haloprof}
\end{center}
\end{table}

\section{EGRET Excess as Mapped by GLAST}
\label{egretglast}     

Much more information on the nature of the EGRET excess at the Galactic
center will be provided by the GLAST telescope. With respect to EGRET, 
GLAST will cover a wider energy range, with an increased effective area and 
better energy and angular resolution. Besides pinning down features in the 
WIMP-induced flux mediated by $\pi^0$ decays, GLAST will have the power to 
search for the monochromatic gamma-ray flux eventually arising from the pair 
annihilation, at 1-loop level, of non-relativistic WIMP's into a two-body final 
state containing a photon (for neutralinos as WIMP dark matter 
candidates, two such states are allowed: 
$\chi\chi \rightarrow 2\gamma$~\cite{bu} and 
$\chi\chi \rightarrow \gamma Z$~\cite{ub}, producing photons with energy
equal to, respectively, $M_\chi$ and $M_\chi~(1-M_z^2/4M_\chi^2)$). 
The discussion of GLAST potential to detect the monochromatic components is 
postponed to a future analysis; we focus here on the term with continuum 
energy spectrum.

We start by supposing that the GC excess as mapped
by EGRET is indeed due to WIMP annihilations and extrapolate what kind
of data GLAST would collect about it. For the performance of the detector, we 
rely on a simplified picture emerging from the latest 
simulations~\cite{glast}. We assume that, on average in the energy interval 
of interest to us, the instrument has an energy resolution of 10\%, 
angular resolution of 10$^{-5}$ sr ($\sim$0.1$^{\circ}$), and a peak 
effective area  of  11000 cm$^2$. GLAST will perform an all sky survey,
rather than operating in the pointing mode. We assume a data acquisition time 
of 2 years and derive the fraction of time the GC center is visible by 
the instrument. We simulate a sky survey with $\pm$ 35$^\circ$ rocking and
take into account the loss of exposure due to the South Atlantic Anomaly (SAA) 
passages, which corresponds to about 14.2\% of the orbiting time. We find that 
the fraction of time that the GC is within the GLAST field of view is
equal to 0.592, and that the net fraction of time that the source 
can be observed is 0.508; the reduction in effective area for sources 
which are not located at the instrument zenith gives a mean effective area 
equal to 60\% of the peak effective area. 

\begin{figure}
\begin{center}
\includegraphics[scale=0.36]{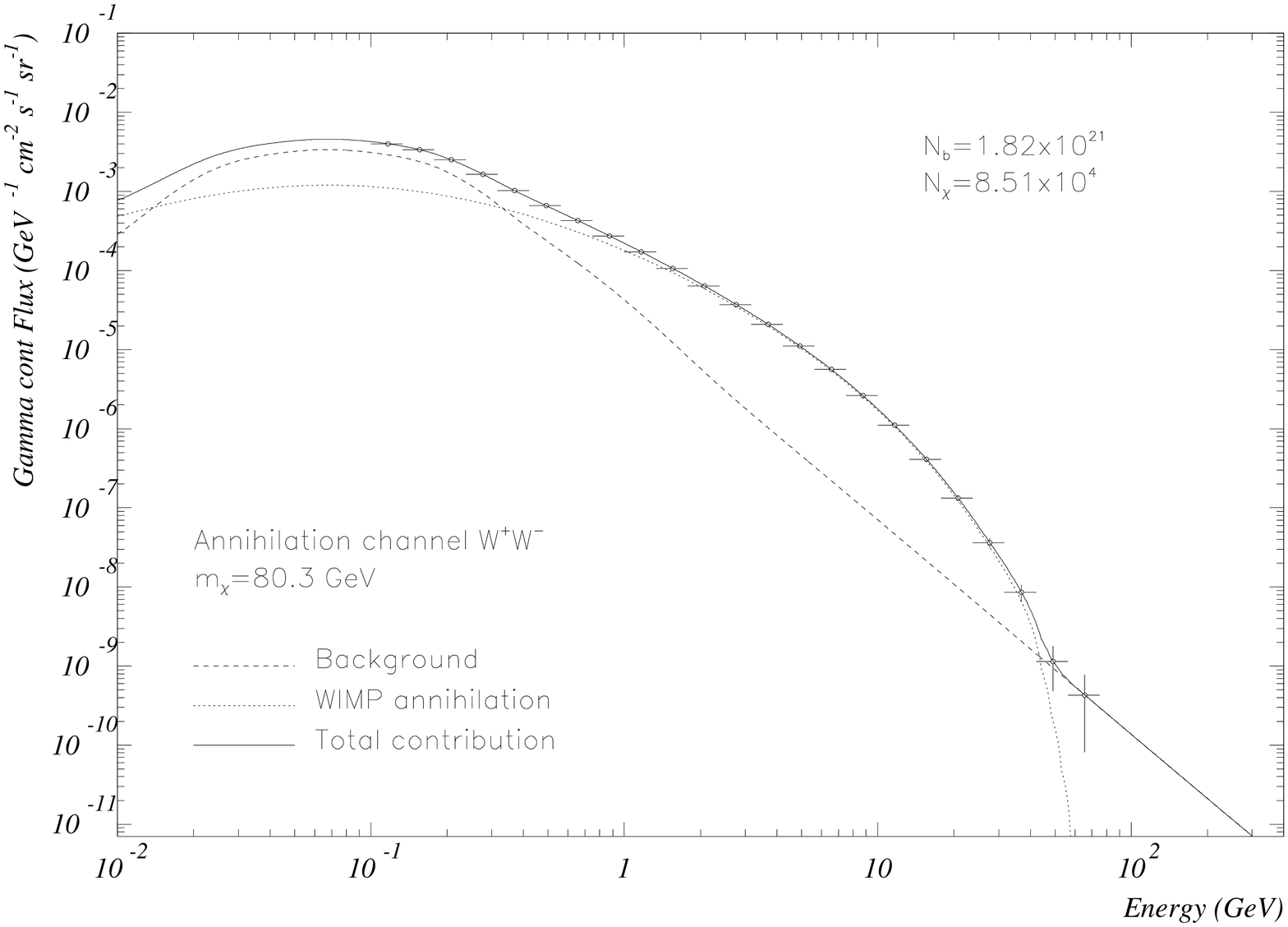}
\caption{Simulation of the data set which will be obtained with GLAST in 
2 years, in case the EGRET GC excess is due to the WIMP-induced flux shown
in one of the sample fits in Fig.~\ref{egretfit} (lower panel). The error bars 
refer to statistical errors for the chosen energy binning and for the angular 
acceptance $\Delta\Omega=10^{-3}\;\rm{sr}$.}
\label{GCglast}
\end{center}
\end{figure}

Fig.~\ref{GCglast} shows a projection for the GC flux which could be measured
by GLAST,  assuming the spectrum and normalization for the dark 
matter source and the normalization for the background as derived from
the fit of the EGRET data in Fig.~\ref{egretfit} (lower panel). The simulated 
data points are derived by assuming as angular acceptance the EGRET angular 
resolution ($\Delta\Omega\sim 10^{-3}$~sr), 
and choosing the energy bin widths to be of the order of  $10\%$  of their central values 
to take advantage of the GLAST energy resolution. 
The error bars displayed are associated to the statistical error only. 
If we try to fit the simulated data with the spectral 
shape of the background and a free normalization, the reduced $\chi^2$ we 
get is higher than $2\cdot 10^3$, much larger than 
the reduced $\chi^2$ we obtained in the corresponding fit for the 
EGRET data set.

GLAST will also allow  to search for an eventual
angular signature of the dark matter source. 
A 10$^{-5}$ sr angular 
resolution implies that the telescope will map the GC resolving 
regions with a precision of 7~pc (assuming the sun galactocentric distance is 8.5~kpc). 

Most models for the distribution of dark matter in the Galaxy, such as
the NFW and the Moore et al. models,
predict an enhancement in the dark matter distribution toward the
Galactic center on a scale which exceeds this size.
There is then the 
chance that WIMP annihilations in the GC region may produce a detectable flux on an 
angular scale exceeding the angular resolution of GLAST.
It is then sensible to investigate whether GLAST will detect 
the flux in the example shown in the Fig.~\ref{GCglast}
as coming from a point source located at the GC or from a 
diffuse source with degrading intensity increasing the angle between
the line of sight and the GC direction. To do that we need to focus
on a specific model for the dark matter density profile in the inner Galaxy.
Inspired by the NFW and the Moore et al. profiles, we assume that  
the WIMP density toward the GC scales like:
\begin{equation}
  \rho(r)=\left\{
  \begin{array}{lc}
    \rho_0\left(\frac{r}{r_0}\right)^{-\gamma},\quad r>r_{min} \\
    \rho_0\left(\frac{r_{min}}{r_0}\right)^{-\gamma},\quad r\le r_{min}
  \end{array}
  \right..
\label{halo}
\end{equation}
We normalize $\rho(r)$ by fixing the local WIMP density, 
i.e. the density at the galactocentric distance $r_0=8.5$ kpc, to be equal to
$\rho_0=0.3$~GeV/cm$^3$. $\gamma$ is kept as a free parameter. To avoid the 
singularity in $r=0$ we have introduced a lower cut-off 
$r_{min}=10^{-5}$~kpc, corresponding to a distance from the GC below which 
we assume that the power law behavior cannot be trusted.

\begin{figure}
\begin{center}
\includegraphics[scale=0.36]{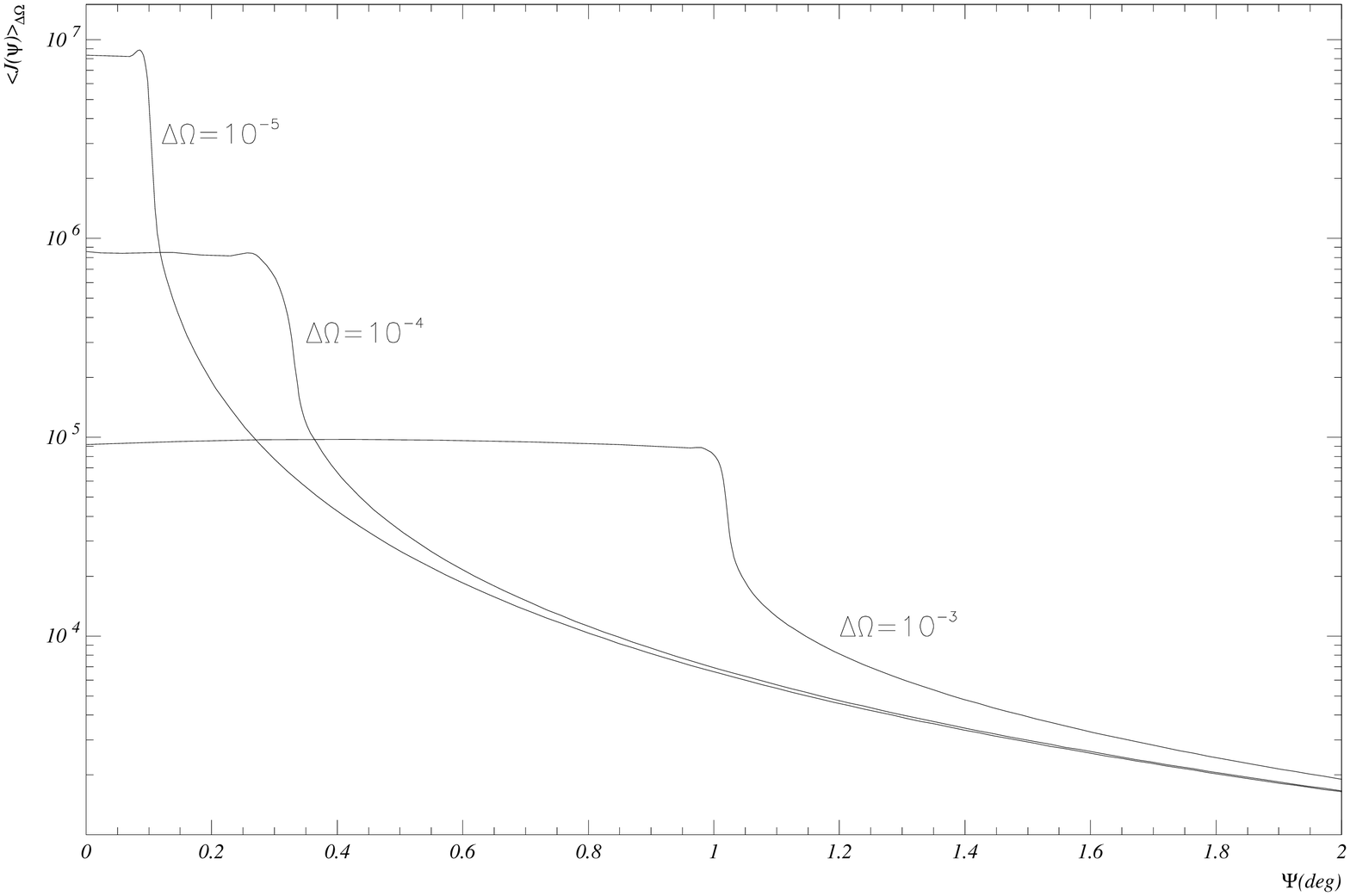}
\caption{Angular dependence for the WIMP signal displayed in the lower panel of
Fig.~\ref{egretfit} and in Fig.~\ref{GCglast} in the case in which the dark matter density
profile $\rho(r)$ has the power law form introduced in Eq.(\ref{halo}).
$\psi$ is the angle between the direction of observation and that of the GC.
$\langle J(\psi) \rangle_{\Delta\Omega}$ coincides with $N_\chi$ for the
toy-model we introduced. Three sample angular acceptances $\Delta\Omega$ are 
considered.}
\label{jpsi}
\end{center}
\end{figure}

\begin{figure}
\begin{center}
\includegraphics[scale=0.36]{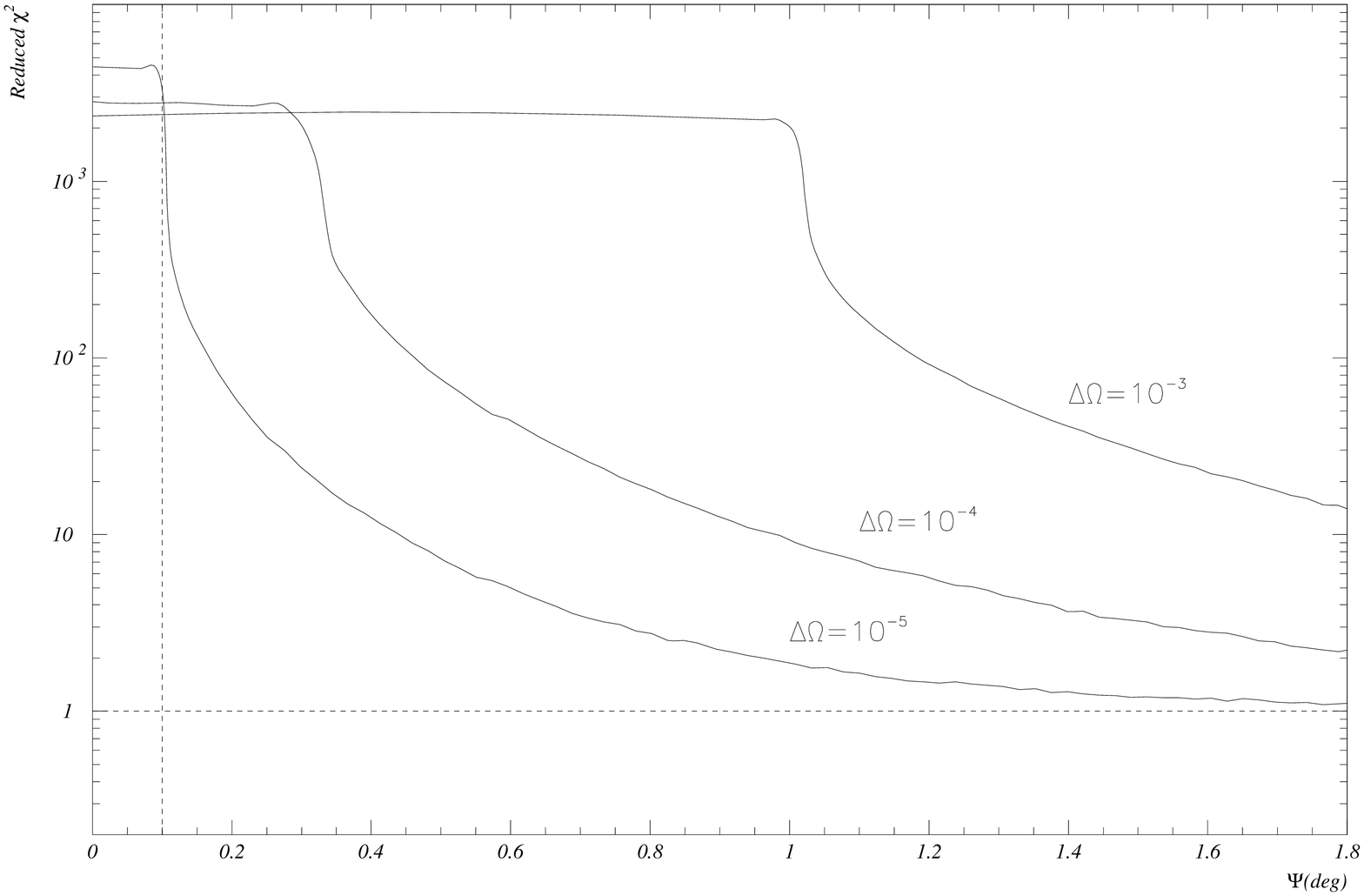}
\caption{For the WIMP-induced flux corresponding to the model shown in Fig.~\ref{GCglast}
and Fig.~\ref{jpsi}, 
we compute the expected flux that would be obtained by GLAST
in 2 years. We fix the angular acceptance $\Delta\Omega$ and denote the angle between
the direction of observation and the GC by $\psi$. 
In the figure we plot the reduced 
$\chi^2$ obtained by fitting the expected flux with a background
component only. We find that $\chi^2\gg 1$ for $\psi$ much
larger than the angular resolution of GLAST (vertical dotted line in the figure).
We then conclude that GLAST will be able to resolve the angular structure of the 
signal for the case considered.}
\label{chi2omega}
\end{center}
\end{figure}

For the sample toy-model shown in Fig.~\ref{egretfit} (lower panel) and in 
Fig.~\ref{GCglast}, $N_\chi = \langle J(0) \rangle_{\Delta\Omega}
(\Delta\Omega = 10^{-3}\;\rm{sr}) = 8.5 \cdot 10^4$ from which we infer that $\gamma = 1.54$. 
Values of $\langle J(\psi) \rangle_{\Delta\Omega}$, for 
such $\gamma$ and for the dark matter density profile specified in Eq.(\ref{halo}), 
are shown in Fig.~\ref{jpsi} as a function of $\psi$ and for a few values of
$\Delta\Omega$.
We can then calculate the expected flux for the model obtained in the fit of EGRET 
data, fixing the direction of observation and the angular acceptance. Finally 
we compute the expected flux that, with the above provisions,
GLAST will collect. We have also included a background
component independent from $\psi$ which is equal to that we have estimated from the fit of 
the data by EGRET.

In Fig.~\ref{chi2omega},
as a function of the direction  $\psi$  and for a few values of
$\Delta\Omega$, we plot the value of the reduced $\chi^2$ we get by fitting in each
case the expected flux with a component that has the spectral shape of the 
background and free normalization. As $\chi^2\gg 1$ on
angular scales much larger than the angular resolution, this indicates that 
GLAST will resolve the angular structure of the dark matter source at the GC. 

>From the figure we see also that, for 
$\psi=0$, the largest $\chi^2$ are obtained for the minimum $\Delta\Omega$ 
considered, i.e. an angular acceptance equal to the angular resolution
of the detector. This indicates that, for the {\em specific} halo 
model considered, the ratio of the signal to square root of the background 
increases going to smaller and smaller angular acceptances.

Values of the reduced $\chi^2$ have been obtained so far by supposing that the
spectral shape of the background is known. Actually, what we have done is to fix
it according to one of the currently favored scenarios.
Slight discrepancies with respect to this model are plausible. On the other
hand, GLAST will perform an all sky survey in which the background component
will be accurately measured at all longitudes and latitudes.
It might still be problematic to choose the background normalization in the 
Galactic center direction, if an excess is indeed found in that direction. 
But it will be possible to relax the assumption 
that the spectral shape of the background is known from theory, as it will 
be possible to extrapolate it from the data at higher latitudes and longitudes. 
A (measured) spectral shape which is different from what we assumed would slightly
change the numerical predictions we derived so far; on the other hand the
general features would remain exactly the same. Keeping this in mind, one should 
not take a reduced $\chi^2$ of 1 as a strict discriminator, e.g. in 
Fig.~\ref{chi2omega}, to decide whether the signal could be resolved from the 
background.

\section{GLAST Performance for a Weaker Source}
\label{glastweak}

Although we have shown that the flux measured by EGRET is compatible with
being due to WIMP annihilations, until more accurate data are available
it will not be possible to discriminate this solution from other plausible
scenarios. In particular, given the rather poor angular resolution of the 
EGRET instrument, there is even the possibility that the EGRET excess is not
actually associated to a source located at the Galactic center. 
This is the case, e.g., if the flux is identified with inverse Compton emission 
from the electrons responsible for the synchrotron emission in the radio arc 
around the Galactic center~\cite{pohlradio}. Should the next generation of
gamma-ray observations solve the puzzle in this direction, there still would
be room for finding a component due to WIMP annihilations in the Galactic 
center region, maybe associated to a dark matter source weaker than the one
we postulated so far. Hence we extend the analysis we performed to investigate 
the potential of GLAST to single out such a source.

As we have shown, for a halo profile of the type given in Eq.(\ref{halo}), it is 
advantageous to focus on a region which is as small 
as possible around the GC. Hence we consider a survey with angular acceptance equal to
the angular resolution of GLAST, $\sim 10^{-5}$~sr. We assume that the background
component is still due to diffuse emission from primary cosmic rays interacting with
the interstellar medium. Hence we keep the spectral form implemented so far 
for $S_b$, with a
normalization such that it matches at least the higher latitude measured flux, i.e.
$N_b > 3.2\cdot 10^{20}$ for a background at least at the level of the flux reported in
the second column in Table~\ref{egretdata}. 
For a given WIMP model, i.e. fixing $S_\chi$, we search then for the minimum ratio
between the two normalization factors $N_\chi/N_b$ that is needed to eventually
discriminate with the GLAST telescope the WIMP annihilation signal from the background.
$N_\chi$ is now equal to $\langle J(0) \rangle_{\Delta\Omega} 
(\Delta\Omega = 10^{-5}\;\rm{sr})$: sample values for this quantity are listed in the
third column of Table~\ref{haloprof} for the three halo profiles introduced in 
Section~\ref{fitsub}. Again, the huge spread in the predictions (possibly even further 
amplified by the redistribution of dark matter particles during the formation of the 
central black hole~\cite{gs,bhullio}) reflects our lack of knowledge about the dark 
matter density in the Galactic center region.

For each pair $N_\chi$ and $N_b$, assuming the same instrumental performance,
exposure time and energy binning specified in Section~\ref{egretglast}, 
we simulate the corresponding data set 
that GLAST would obtain, i.e. the expected flux measurements with the associated 
statistical error for the chosen energy binning, analogously to what is shown in 
Fig.~\ref{GCglast} for the EGRET GC source. The criterion to discriminate whether
the WIMP component would be singled out is based on the usual $\chi^2$
test statistic. We have computed the reduced
$\chi^2$ between the number of counts expected in each energy bin for the
two hypothesis: WIMP signal plus background and background
only. Taking into account the number of degrees of freedom, which in our
case is equal to the number of energy bins, the signal plus background
curve is distinguishable from the background only curve, for a reduced
$\chi^2>constant$. This constant in uniquely determined by the number of
degrees of freedom and by the confidence level we want to reach.  We have
also checked our results against those obtained with the likelihood ratio 
method \cite{hu-nielsen,read}, obtaining no 
discrepancies.\footnote{We thank G.Ganis for providing the software package 
we used to perform this analysis.} 
This latter method is especially suited for the case we have at hand:
to decide whether a certain
event belongs to the background only hypothesis ($H_0$) or to signal plus background 
hypothesis ($H_1$), one starts by constructing two probability distributions, $P_0$ and 
$P_1$, for an estimator $F={ L}(H_1)/{ L}(H_0)$, which is the ratio between 
the likelihoods ${ L}$ of the two hypotheses.
In our case, since we are interested in counting, we can choose the 
Poisson distribution to obtain the likelihood. Comparing the
two distributions one can decide, at a certain confidence level, if they
will result distinguishable or not, once it is fixed the accuracy of the
experimental data that will be used for the discrimination.
The
likelihood ratio method is in general more powerful than the $\chi^2$
one, since, in addition to giving the probability of a certain set of
data to belong to the signal plus background probability distribution, it
allows to compute the probability to be wrong when accepting such
hypothesis, the so called {\it power of the test}, considering the background only 
hypothesis as the \emph{true} one.

\begin{figure}[t]
\begin{center}
\includegraphics[scale=0.36]{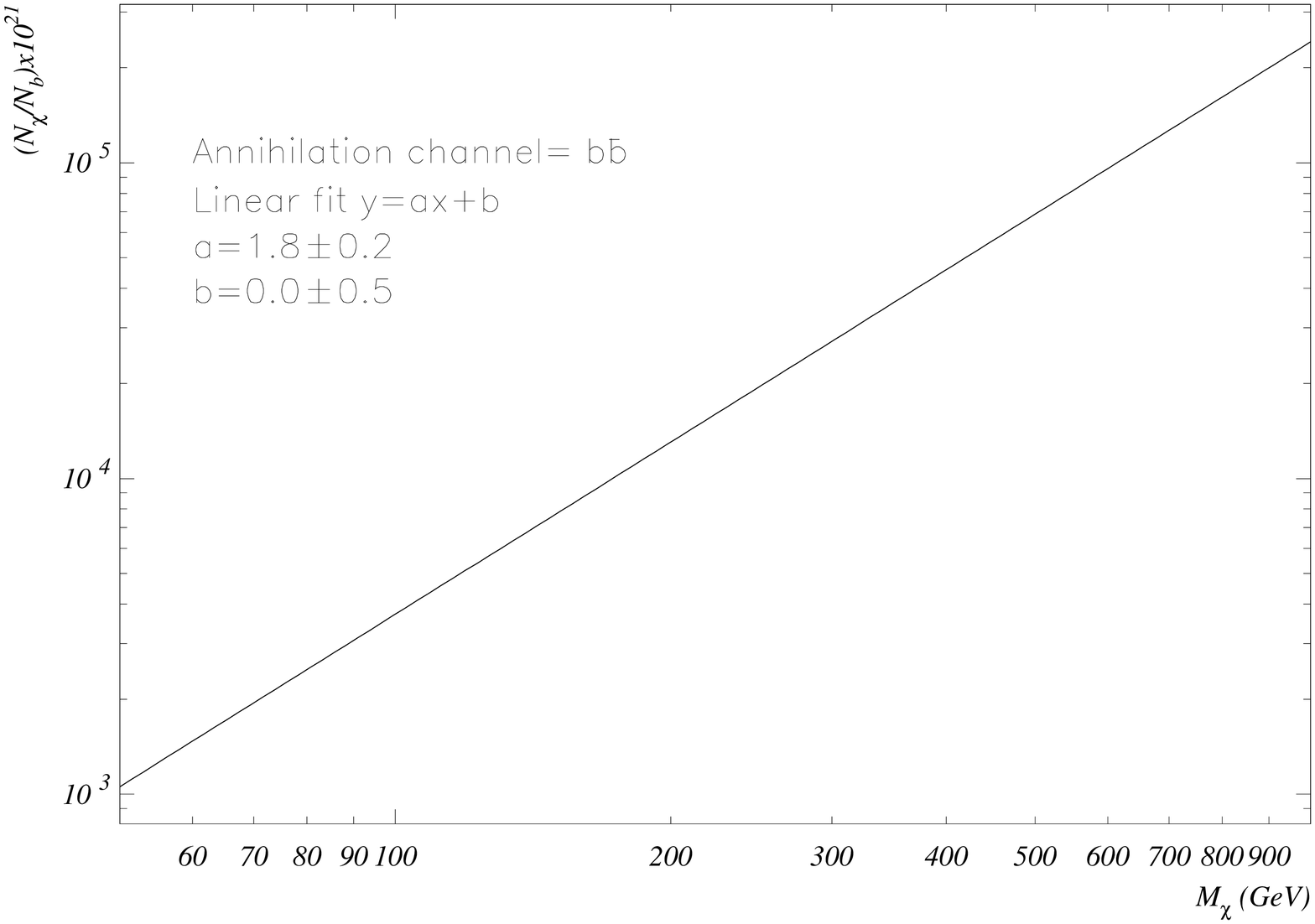}
\caption{Minimum ratio between the normalization of the WIMP signal $N_\chi$
and the background normalization $N_b$ such that the WIMP induced signal would be
singled out of the background with GLAST. We are referring to a toy-model with
a single annihilation channel allowed, i.e. $b\bar{b}$ in the case displayed.
In the linear regression fit marked in the figure 
$y=\log_{10}\left(\frac{N_\chi}{N_b}\right)$ and $x=\log_{10}M_\chi$.}
\label{glastweakfig}
\end{center}
\end{figure}

As a sample test case, we consider again the toy-model introduced in 
Section~\ref{fitsub} for a single WIMP annihilation channel, e.g., $b\bar{b}$.
In Fig.~\ref{glastweakfig} we plot, as a function of the WIMP mass, the minimum value 
of the ratio $N_\chi/N_b$  needed for a discrimination of the WIMP signal from
the background. The corresponding values of $N_\chi = \langle J(0) \rangle_{\Delta\Omega} 
(\Delta\Omega = 10^{-5}\;\rm{sr})$ are of the order of $\sim 10^3$ for a WIMP mass
of about 50~GeV. They increase approximately as $M_\chi^{1.8}$ for heavier particles.
These values are larger than the ones one would obtain from a 
smooth profile, see Table~\ref{haloprof}. 
A local enhancement in the WIMP dark matter density at the GC is then required to match 
our values of $N_\chi$.
Such enhancement seems to be smaller than the one needed to fit the EGRET excess in the previous section 
($N_\chi$ of the order of $\sim 10^4\div 10^5$). 
As a word of caution we remark that the values of $N_\chi$ we obtained in the two analysis 
should not be directly compared in principle
as the angular acceptance in the two cases is different: $\Delta\Omega = 10^{-5}\;\rm{sr}$
in the current case, while we had $\Delta\Omega = 10^{-3}\;\rm{sr}$ when we fitted
the EGRET data set. A halo model has to be specified to translate one into the other.

\section{A Specific WIMP: the Lightest Neutralino in the mSUGRA Framework} 
\label{sugra}

As a sample application of the generic tool we discussed so far, we focus now
on the most widely studied WIMP dark matter candidate, the lightest neutralino, 
in the most restrictive supersymmetric extension of the Standard Model,
the minimal supergravity (mSUGRA) framework~\cite{msugra}. 
This setup has been considered extensively in the contest of dark matter detection 
(a list of recent references includes, e.g.,~\cite{feng,msd5,msd4,msd3,msd2,msd1})
and therefore the comparison of our results
with previous work and other complementary techniques should be transparent in this 
case.

In the general framework of the minimal supersymmetric extension of the Standard Model
(MSSM), the lightest neutralino is the lightest mass eigenstate obtained from the 
superposition of four interaction eigenstates, the supersymmetric partners of the neutral 
gauge bosons (the bino and the wino) and Higgs bosons (two Higgsinos). 
Its mass, composition and couplings with Standard Model particles 
and other superpartners are a function of the several free parameters one needs to 
introduce to define such supersymmetric extension. In the mSUGRA model, universality 
at the grand unification scale is imposed. With this assumption the number of free
parameters is limited to five: 
\[ m_{1/2},\;\;  m_0,\;\;  sign(\mu),\;\;  A_0\;\; \rm{and}\;\; \tan\beta \;,
\]
where $m_0$ is the common scalar mass, $m_{1/2}$ is the common gaugino mass and $A_0$ 
is the 
proportionality factor between the supersymmetry breaking trilinear couplings and the 
Yukawa couplings. $\tan\beta$ denotes the ratio of the vacuum
expectation values of the two neutral components of the SU(2) Higgs doublet, while
the Higgs mixing $\mu$ is determined (up to a sign) by imposing the Electro-Weak Symmetry 
Breaking (EWSB) conditions at the weak scale. In this context 
the MSSM can be regarded as an effective low energy theory. The parameters at the weak energy scale 
are determined by the evolution of those at the unification scale, according to the renormalization 
group equations (RGEs). For this purpose, we have made use of the ISASUGRA RGE package 
in the ISAJET 7.64 software~\cite{isasugra}.

After fixing the five mSUGRA parameters at the unification scale, we extract from the ISASUGRA output
the weak-scale supersymmetric mass spectrum and the relative mixings. Cases in which the
lightest neutralino is not the lightest supersymmetric particle or there is no
radiative EWSB  are disregarded.
The ISASUGRA output is then used as an input in the \ds\ package\cite{ds}. The latter
is exploited to: 
\begin{itemize}
\item reject models which violate limits recommended
by the Particle Data Group 2002 (PDG) \cite{pdg02};
\item compute the neutralino
relic abundance, with full numerical solution of the density evolution equation
including resonances, threshold effects and all possible coannihilation 
processes~\cite{newcoann}; 
\item compute the neutralino annihilation rate at zero 
temperature in all kinematically allowed tree-level final states (including fermions,
gauge bosons and Higgs bosons;
\item estimate the induced gamma-ray yield by 
linking to the results of the simulations performed with the Lund Monte Carlo 
program Pythia~\cite{pythia} as implemented in the \ds package.
\end{itemize}

Note that none of the approximations  implemented in the toy-model introduced above,
regarding the estimate  of the relic density or the annihilation cross section,
are applied here.

\begin{figure}[t]
\begin{center}
\includegraphics[scale=0.50]{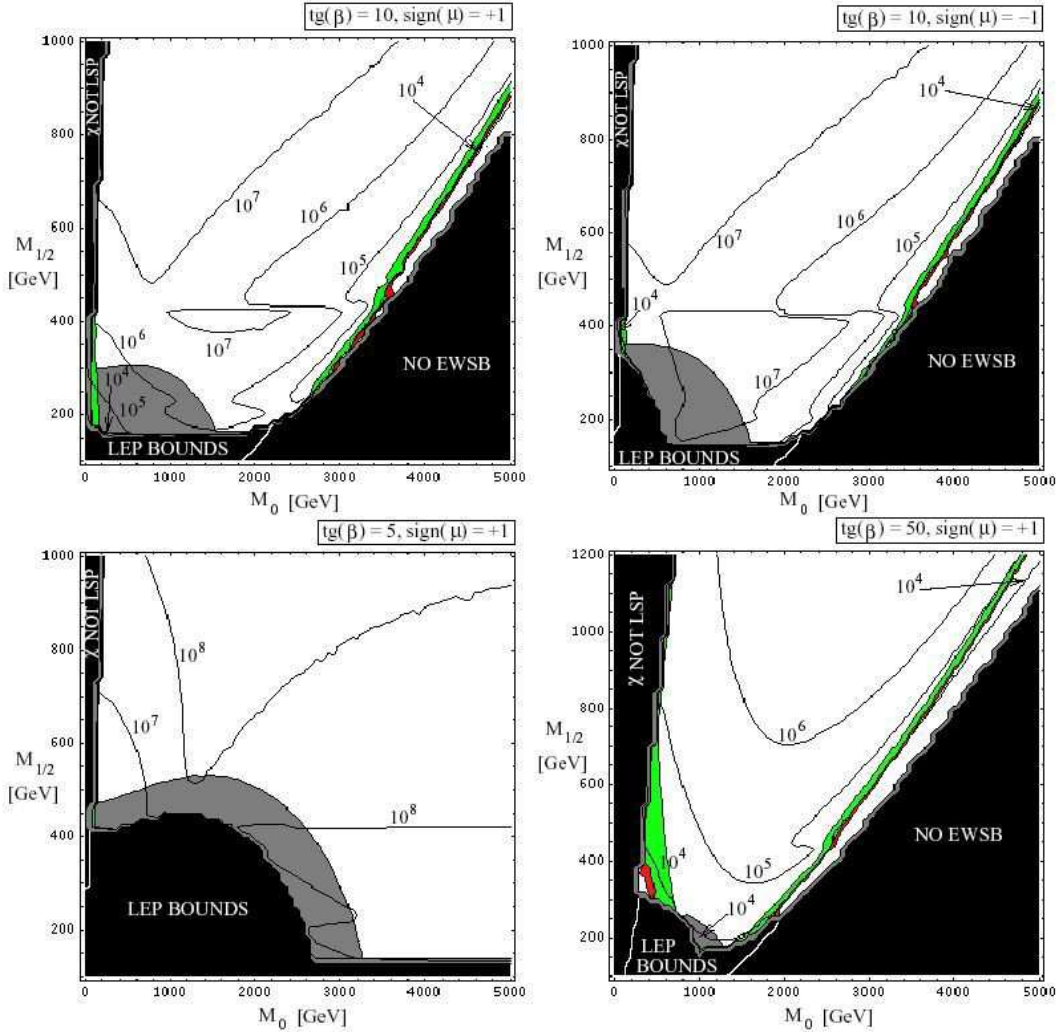}
\caption{Contour plots in the mSUGRA $(m_0,m_{1/2})$ plane, for the value
of the normalization factor $N_\chi$, that allows the detection of the
neutralino $\gamma$ ray signal, with GLAST. In the green region 
$0.13\le\Omega_{\chi} h^2\le 0.3$, while the red region corresponds to the WMAP range $0.09\le\Omega_{\chi} h^2\le 0.13$~\cite{wmap}. The black region corresponds to models that are excluded either by incorrect EWSB, LEP bounds violations~\cite{pdg02} or because the neutralino is not the LSP. In the dark shaded region $m_{h_0}<114.3$ GeV~\cite{pdg02}, where $h_0$ is the lightest Higgs.}
\label{fig4}
\end{center}
\end{figure}

\begin{figure}[ht]
\begin{center}
\includegraphics[scale=0.50]{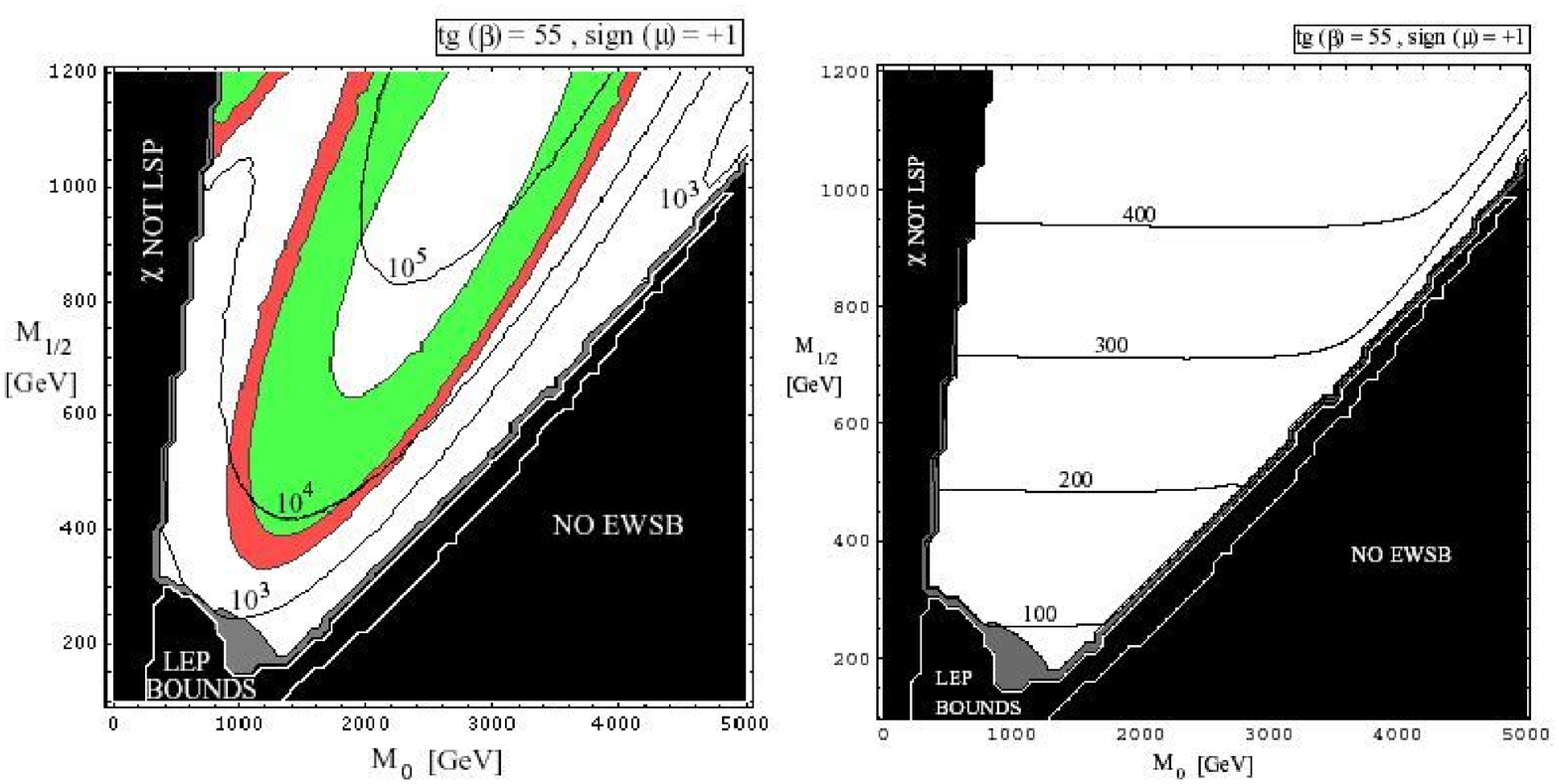}
\caption{Contour plots in the mSUGRA $(m_0,m_{1/2})$ plane for $\tan\beta =55$. Left panel: values
of the normalization factor $N_\chi$, that allow the detection of the
neutralino $\gamma$ ray signal, with GLAST. 
Right panel: values of the neutralino mass.
The excluded and colored regions are as in Fig.~\ref{fig4}.}
\label{figtb55}
\end{center}
\end{figure}

We are ready then to exploit the procedure outlined in Section~\ref{glastweak}
to study in what region of the mSUGRA parameter space, for a given dark matter halo 
profile, the induced continuum $\gamma-$ray flux would be detectable 
by GLAST. Fixing $\tan\beta$, $A_0$ and $sgn(\mu)$, we have performed a scan in the 
$(m_0,m_{1/2})$ plane searching for the minimum dark matter density, in the GC
region, needed to be able to single out the neutralino annihilation signal with GLAST.
To do this we have followed the same discrimination criteria described in 
Section \ref{glastweak} that we recapitulate here for the reader's benefit.
First we estimate the statistical error (1$\sigma$) on GLAST data to be the square 
root of the number of 
events. To compute the latter we multiply the flux by the effective area of the detector, by the total
observational time and the angular resolution $\Delta\Omega = 10^{-5}\;\rm{sr}$.
Then for each value of the pair $(m_0,m_{1/2})$ we compute the difference between the fluxes
$\phi_\gamma= \phi_b + \phi_\chi = N_b { S}_b + N_\chi { S}_\chi$
and $\phi^\prime_\gamma= \phi_b=  N_b { S}_b$. If $\phi_\gamma-\phi^\prime_\gamma > 3\sigma$
we consider the SUSY model with those values of $(m_0,m_{1/2})$ to be detectable by GLAST.
In Figs.~\ref{fig4} and~\ref{figtb55} we show the isolevel curves 
for the minimum allowed value of $ N_\chi=\langle J(0) \rangle_{\Delta\Omega} 
(\Delta\Omega = 10^{-5}\;\rm{sr})$ for the signal detection, in the $(m_0,m_{1/2})$
plane and for five sample sets of the other parameters.

The colored regions in the figures represent portions of the parameter space where the neutralino has the right cosmological relic density to constitute CDM: in the green region $0.13\le\Omega_{\chi} h^2\le 0.3$, while the red region corresponds to the WMAP range $0.09\le\Omega_{\chi} h^2\le 0.13$~\cite{wmap}.
It is interesting to remark that in the cosmologically favored regions the values of $\langle J \rangle$ 
which allow the detection of the WIMP signal by GLAST are among the smallest in the 
interval of variability of the variable $\langle J \rangle$ itself.
To some extent this was anticipated in Eq.(\ref{eq:scal}) that was one of the 
approximations of our toy-model: low values of the relic abundance lead to high
values of the annihilation cross section. 

The cosmologically favored regions correspond to two regimes which have been extensively 
studied in the
literature. The first regime is given by models where the neutralino tends to be a very pure bino.
It comprises the region where the neutralino is light (lower parts of each panel, 
compare with Fig. \ref{fig5}) and the region where the stau coannihilation is active (the low $m_0$ region on the left of the first three panels, see, e.g.,~\cite{ellisstau}).
The second regime,
sometimes dubbed as ``focus-point'' region~\cite{focus1,focus2}, is the region on the 
right-hand side of the panels with $\tan\beta$ equal to 10, 50 and 55, close to
the region in parameter space where there is no EWSB, in which the neutralino
has a relevant Higgsino component.

Comparing the values of $\langle J(0) \rangle_{\Delta\Omega} 
(\Delta\Omega = 10^{-5}\;\rm{sr})$ obtained with those listed in Table~\ref{haloprof},
we see that in the case of a moderate enhancement of the dark matter density toward the
GC as predicted by the NFW profile or an even slightly milder singular profile, there is 
a fair portion of the mSUGRA parameter space which gives fluxes detectable with GLAST.
It is worth to observe that high $\tan\beta$ models (\emph{i.e.} $\tan\beta=50$ and $\tan\beta=55$) possess the largest cosmologically favored regions. Furthermore they give the highest chances to single out the 
$\gamma$-ray signal from neutralino annihilations with GLAST.
We can compare our results with those of~\cite{feng}. There it is assumed that
GLAST has a certain sensitivity to the integrated continuum $\gamma-$ray flux from 
a region around the GC, of an extension $10^2$ times wider than the GLAST angular
resolution. The neutralino-induced signal is then supposed to be
detectable if its integrated flux exceeds such sensitivity. 
In Figs. 18 and 19 of~\cite{feng} are given the regions which can be probed by GLAST in case of 
$\left<J(0)\right>_{\Delta\Omega}(\Delta\Omega= 10^{-3}\;\rm{sr})=500$.
They are in qualitative agreement with  our corresponding predictions (first and fourth panels 
of Fig.~\ref{fig4}) for $\Delta\Omega=10^{-5}$~sr. 

\begin{figure}[t]
\begin{center}
\includegraphics[scale=0.50]{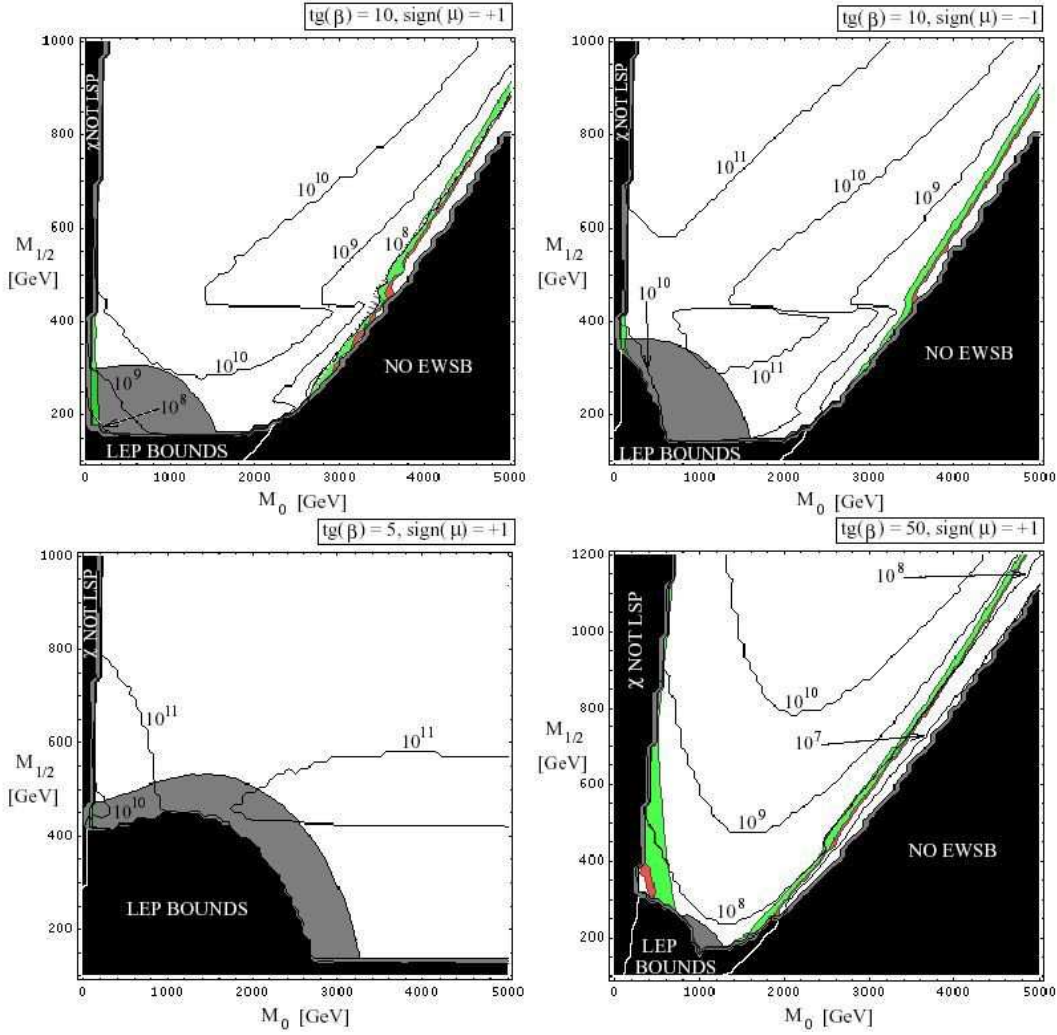}
\caption{Contour plots in the mSUGRA $(m_0,m_{1/2})$ plane, for values of
$N_\chi$ that are already excluded by EGRET data at $5\sigma$ confidence
level. In the green region $0.13\le\Omega_{\chi} h^2\le 0.3$, while the red region corresponds to the WMAP range $0.09\le\Omega_{\chi} h^2\le 0.13$~\cite{wmap}. The black region corresponds to models that are excluded either by incorrect EWSB, LEP bounds violations~\cite{pdg02} or because the neutralino is not the LSP. In the dark shaded region $m_{h_0}<114.3$ GeV~\cite{pdg02}, where $h_0$ is the lightest Higgs.}
\label{fig5a}
\end{center}
\end{figure}

\begin{figure}[ht]
\begin{center}
\includegraphics[scale=0.5]{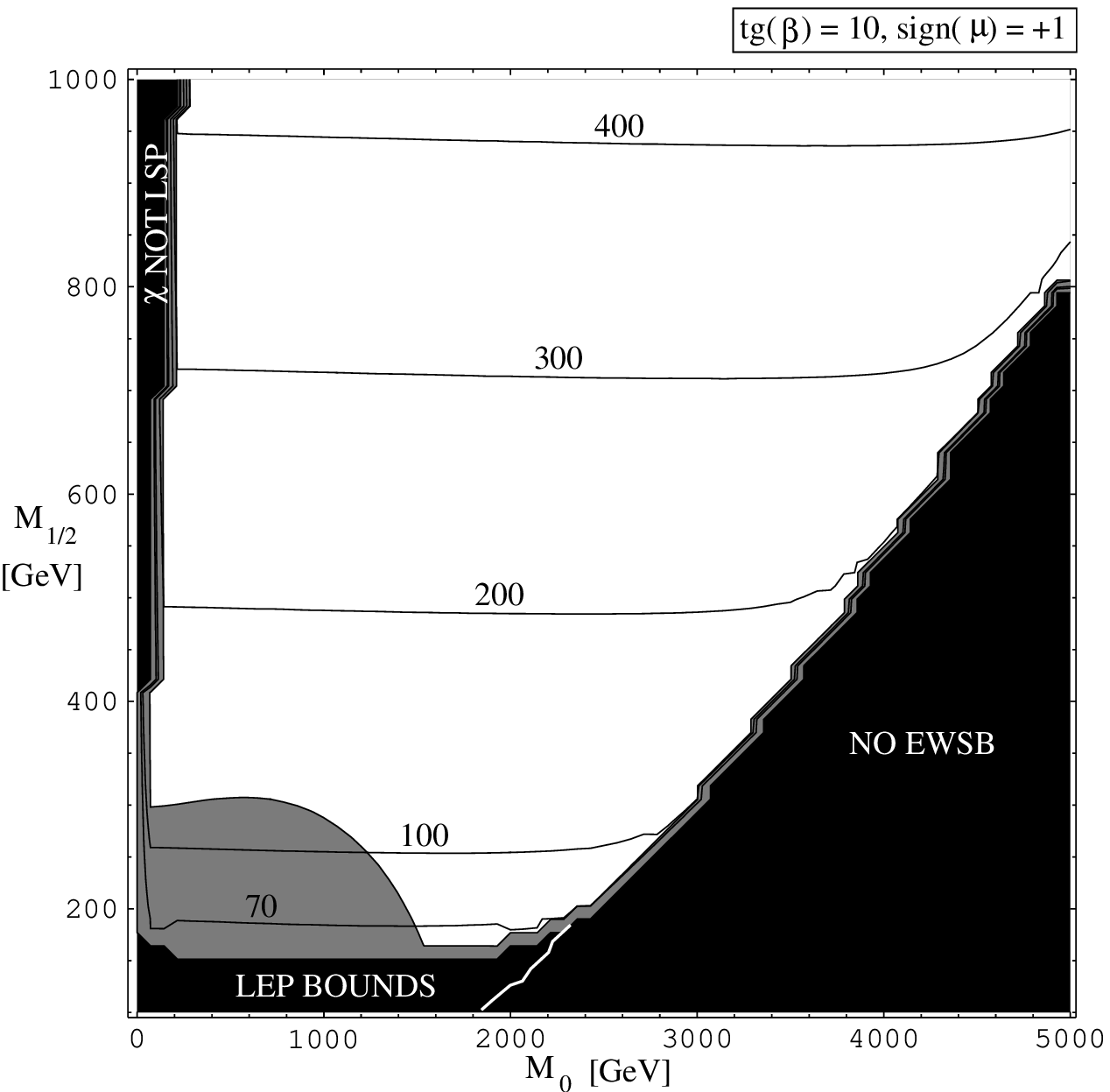}
\includegraphics[scale=0.5]{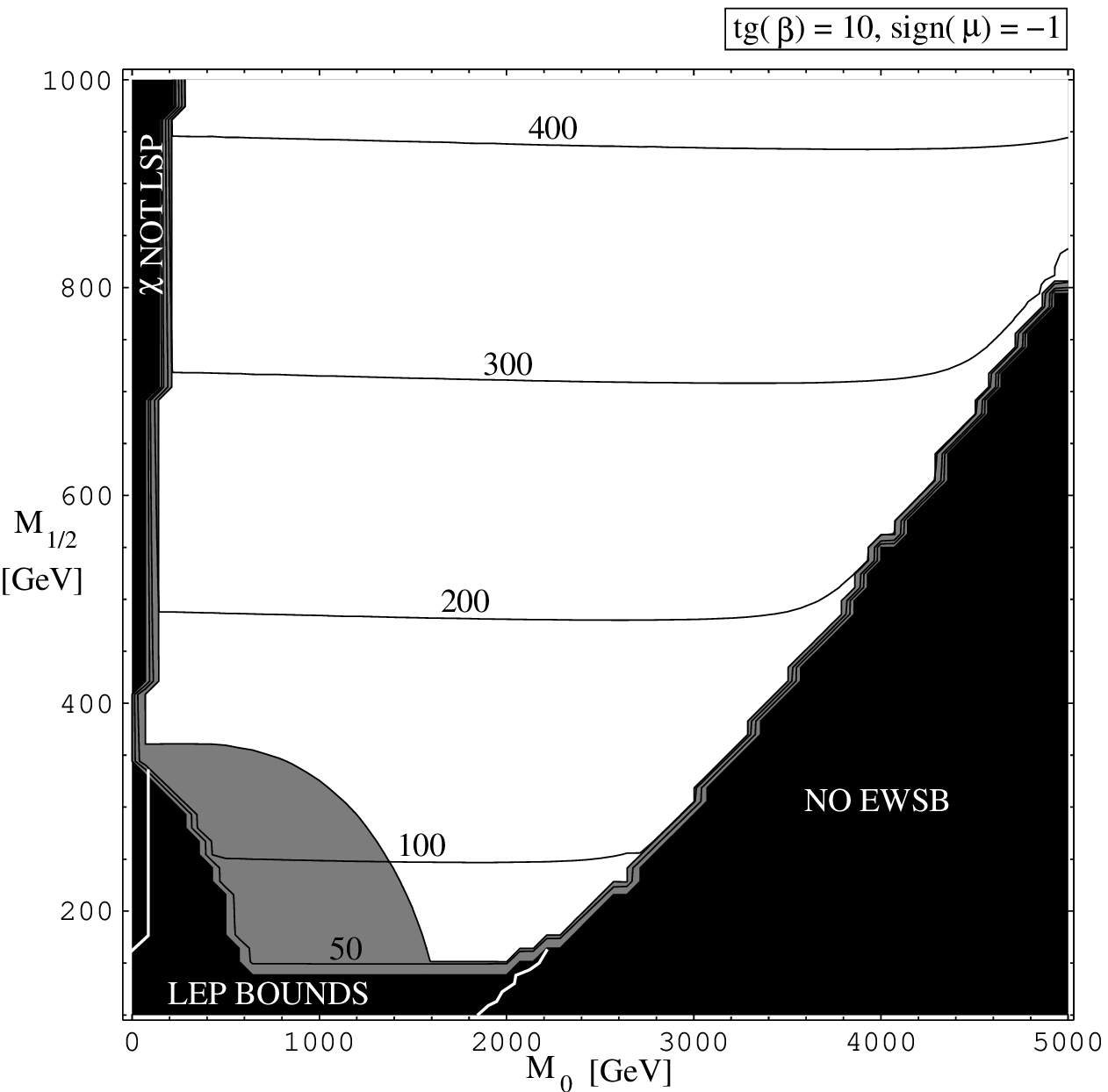}
\includegraphics[scale=0.5]{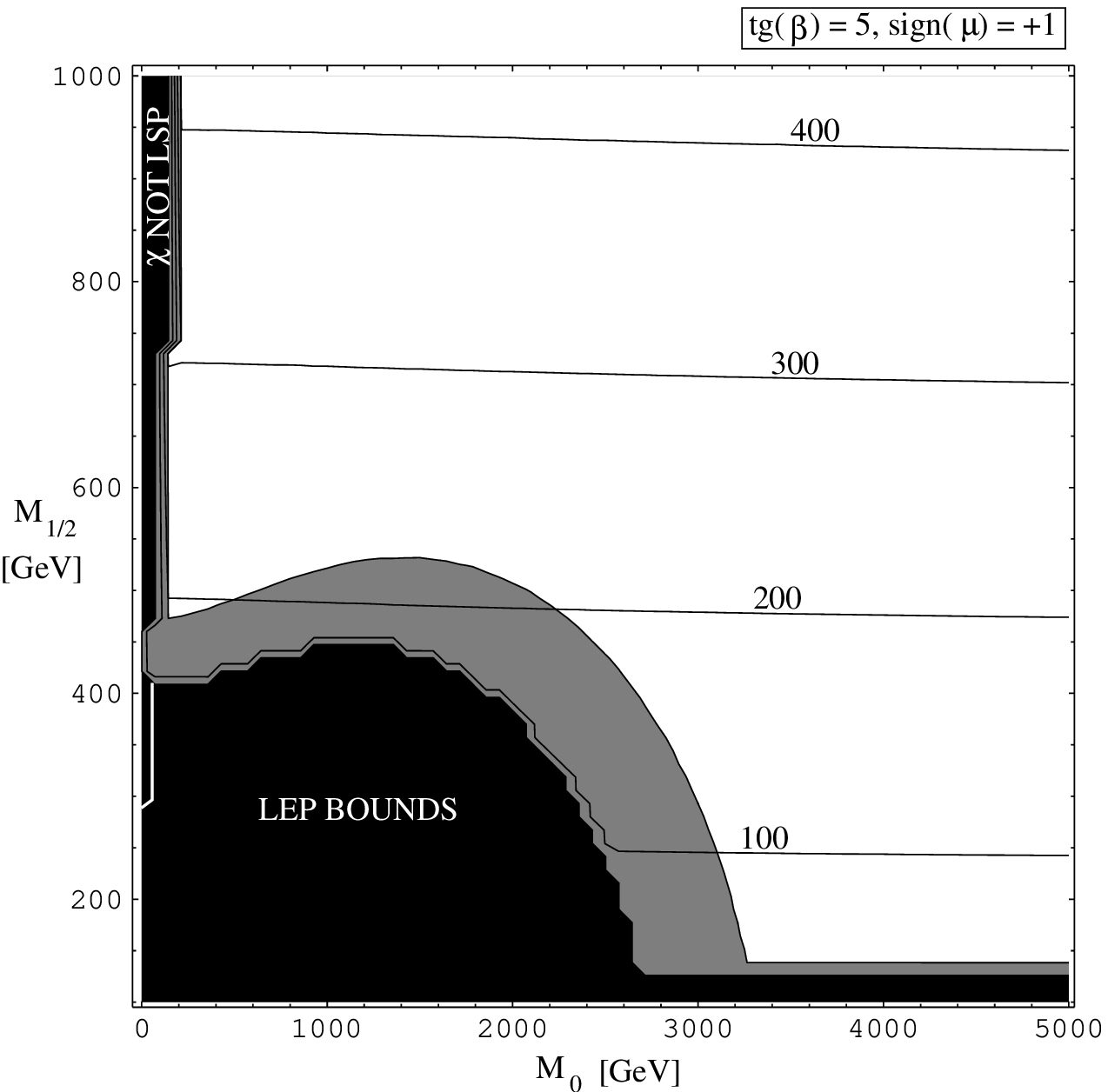}
\includegraphics[scale=0.5]{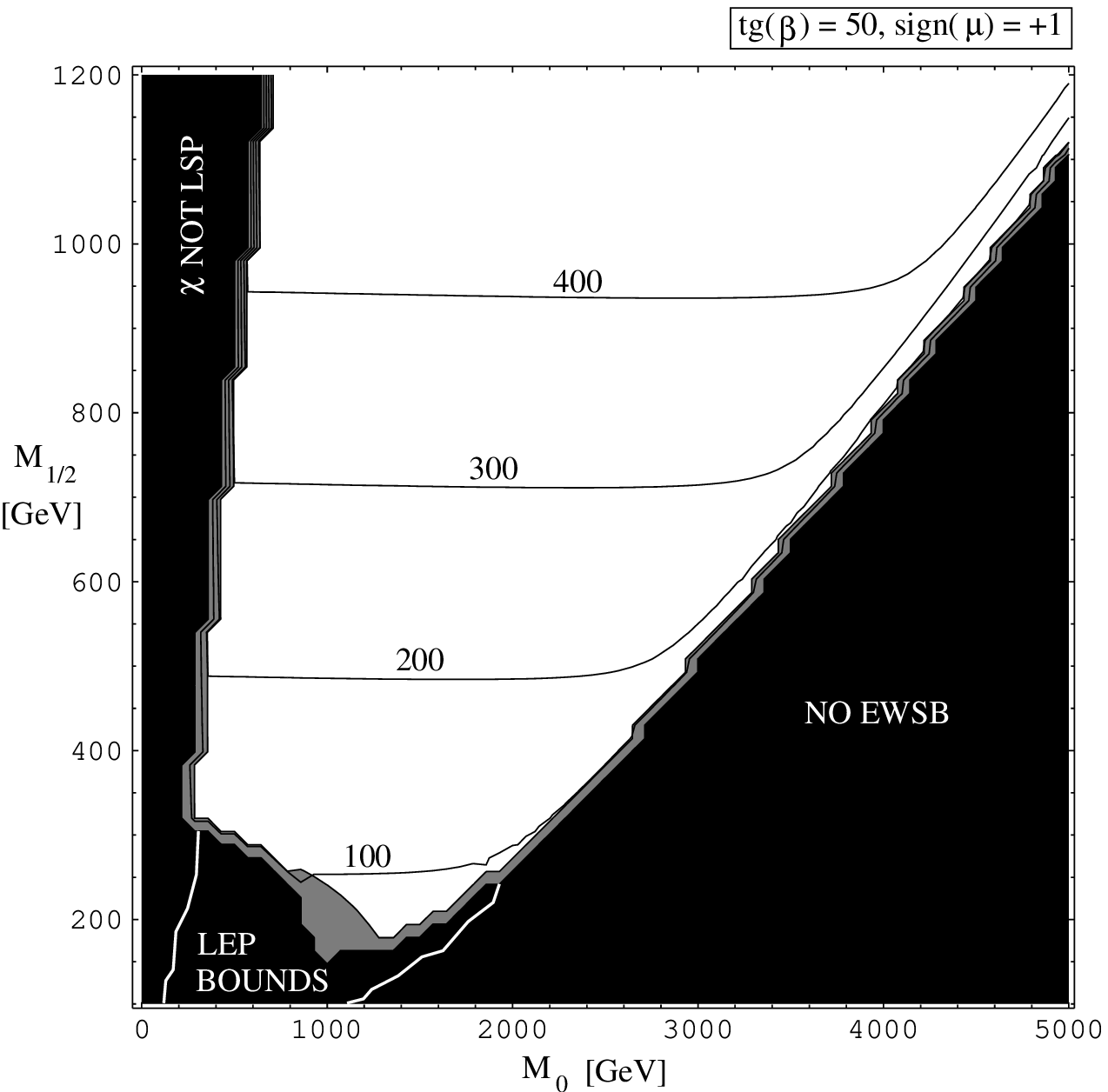}
\caption{Neutralino mass contour plots in the $(m_0,m_{1/2})$ mSUGRA
plane. The black region corresponds to models that are excluded either by incorrect EWSB, LEP bounds violations~\cite{pdg02} or because the neutralino is not the LSP. In the dark shaded region $m_{h_0}<114.3$ GeV~\cite{pdg02}, where $h_0$ is the lightest Higgs.}
\label{fig5}
\end{center}
\end{figure}

In addition to this study of the GLAST sensitivity, we have tried to single out the 
regions of the mSUGRA parameter space ($(m_0,m_{1/2})$ for fixed
$\tan(\beta)$, $A_0$ and $sign(\mu)$), which are already experimentally
excluded, due to a neutralino-induced $\gamma$-ray flux
exceeding the GC EGRET data of table \ref{egretdata}. 
In Fig.~\ref{fig5a} we show isolevel curves for the maximal
$\langle J(0) \rangle_{\Delta\Omega}(\Delta\Omega = 10^{-3}\;\rm{sr})$
marginally consistent with the data at the $5\sigma$ level.
We can observe that the values we obtain are generally significantly larger
than the corresponding sample values in Table~\ref{halo}. This result implies 
that rather weak constraints on the mSUGRA parameter space can actually 
be imposed, on the basis of the GC $\gamma-$ray flux measured by EGRET.

One could finally wonder how the GC EGRET data could be fitted in the context of the mSUGRA models. 
In order to answer to this question, we rely on the general analysis that we have performed in the 
toy-model scheme (see section~\ref{foed}). In particular we could extract from Fig.~\ref{isochi2} the
values of the neutralino mass that allow the best fit of the GC EGRET data\footnote{Recall that the 
possibility to fit EGRET data is quite insensitive to the dominant annihilation channel.}.
Looking at Fig.~\ref{fig5} we can then single out the $(m_0,m_{1/2})$ parameter regions that corresponds to
such values. 

\section{Conclusions}
\label{concl}

We have found that the excess in the $\gamma$-ray flux detected by the EGRET 
telescope toward the Galactic center shows spectral features which are compatible
with an exotic component due to WIMP annihilations, especially for WIMP masses
in the lower end of the mass range currently considered for WIMP dark matter
candidates. For the WIMP-induced flux to be at the level of the measured flux,
a fairly large dark matter density is needed in the Galactic center region;
indeed, such density enhancements are found in N-body simulations of halo profiles
in cold dark matter cosmologies.

Although it is not possible with present data on the Galactic center excess to
discriminate between the interpretation we propose here and other viable explanations,
we have shown that, with the data that will be collected by the GLAST,
the next major $\gamma$-ray telescope in space, it will be possible
to identify both spectral and angular signature expected for a WIMP-induced
component. If on the other hand the data will point to an alternative explanation,
there will still be the chance for the GLAST telescope to single out a (weaker)
dark matter source. The potentials of GLAST have been explored both in the contest
of a generic simplified toy-model for WIMP dark matter and for one of
the most widely studied WIMP dark matter candidate, the lightest neutralino, 
in the minimal supergravity framework. We find that even in case of moderately
singular dark matter profiles, there are regions in the parameter space which
will be probed by GLAST, especially in the high $\tan\beta$ case.
We find, on the contrary, that limits from current 
EGRET data are rather weak.

\section*{Acknowledgments}

We would like to thank all the component of the GLAST Dark Matter working group for
lots of discussion on the subject,  in particular  Elliot Bloom and Eduardo
do Couto e Silva. We also thank Hans Mayer-Hasselwander for providing a table of 
the EGRET data toward the Galactic center and G.Ganis and C.Pittori for many 
useful discussions.


\begin{thebibliography}{0}

\bibitem{wmap}
  D.N.~Spergel et al., astro-ph/0302209.

\bibitem{lars} 
  L. Bergstr{\"o}m, Rept. Prog. Phys. {\bf 63} (2000) 793. 	

\bibitem{silksrednicki} 
  J.~Silk and M.~Srednicki, Phys.\ Rev.\ Lett.\ {\bf 53} (1984) 624.

\bibitem{stecker}
  F.W.~Stecker, S.~Rudaz and T.F.~Walsh, 
  Phys.\ Rev.\ Lett.\ {\bf 55} (1985) 2622.

\bibitem{p1}   
  L.~Bergstr\"om, J.~Edsj\"o and C. Gunnarsson
  Phys. Rev. {\bf D63} (2001) 083515

\bibitem{p2}   
  L.~Bergstr\"om, J.~Edsj\"o and P.~Ullio,
  Phys. Rev. Lett. {\bf 87} (2001) 251301.

\bibitem{p3}   
  G.~Bertone, G.~Sigl and J.~Silk 
  Mon. Not. Roy. Astron. Soc. {\bf 337} (2002) 98.

\bibitem{p4}   
  D.~Merritt, M.~Milosavljevic, L.~Verde and R.~Jimenez,
  Phys. Rev. Lett. {\bf 88} (2002) 191301. 

\bibitem{p5}   
  P.~Ullio, L.~Bergstr\"om, J.~Edsj\"o and C.~Lacey,
  Phys. Rev. {\bf D66} (2002) 123502.

\bibitem{p6}   
  R. Aloisio, P. Blasi and A. V. Olinto, astro-ph/0206036.

\bibitem{p7}   
  D.~Hooper and B.~Dingus, astro-ph/0210617.

\bibitem{Mayer}  
  H.~Mayer-Hasselwander \etal,  Astron. Astrophys. {\bf 335} (1998) 161.

\bibitem{smapj} 
  A.~W.~Strong, I.~V.~Moskalenko and O.~Reimer,
  Astrophys.\ J.\  {\bf 537} (2000) 763
  [Erratum-ibid.\  {\bf 541} (2000) 1109].

\bibitem {glast} 
  Proposal for the Gamma-ray Large Area Space Telescope, SLAC-R-522  (1998);
GLAST Proposal to NASA A0-99-055-03 (1999). 

\bibitem {Bellazzini} 
   R.~Bellazzini, Frascati Physics Series Vol.XXIV, 353, (2002), \\
   (http://www.roma2.infn.it/infn/aldo/ISSS01.html).

\bibitem {morselli} 
   A.Morselli, Frascati Physics Series Vol.XXIV, 363, (2002),\\ 
   (http://www.roma2.infn.it/infn/aldo/ISSS01.html).

\bibitem{Mayerpriv}  
  H.~Mayer-Hasselwander, private communication.


\bibitem{stecker-pi0}
  F.W.~Stecker, Astrophysics and Space Sci.\ {\bfseries 6}, 377 (1970);

\bibitem{gaisserbook}
  T.K.~Gaisser, {\it Cosmic rays and particle physics}, 1990,
  Cambridge University Press, Canmbridge. 

\bibitem{bess}
  BESS Collaboration, T.~Sanuki et al., 
  Astrophys.\ J.\  {\bf 545} (2000) 1135.

\bibitem{ams}
  AMS Collaboration, J.~Alcaraz et al., 
  Phys. Lett. {\bf 472} (2000) 215. 

\bibitem{anomsu}
   P.~Ullio, JHEP {\bf 0106} (2001) 053.

\bibitem{pythia}
  Pythia program package, 
  see T. Sj\"ostrand, Comp.  Phys. Comm. {\bf 82} (1994) 74. 

\bibitem{ds} 
   P.~Gondolo, J.~Edsj\"o, P.~Ullio, L.~Bergstr\"om, M.~Schelke and E.A.~Baltz,
  proceedings of idm2002, York, England, September 2002, astro-ph/0211238; 
  {http://www.physto.se/\~{}edsjo/darksusy/}.

\bibitem{bub}
  L.~Bergstr{\"o}m, P.~Ullio and J.~Buckley, 
  Astropart.\ Phys.\ {\bf 9} (1998) 137.


\bibitem{lee-weinberg}
  B.W.~Lee and S.~Weinberg, Phys.\ Rev.\ Lett.\ {\bfseries 39} (1977) 165.

\bibitem{jkg}
  G.~Jungman, M.~Kamionkowski and K.~Griest,
  Phys. Rep. {\bf 267} (1996) 195.

\bibitem{nfw}
  J.F. Navarro, C.S. Frenk and S.D.M. White,
  Astrophys.\ J.\  {\bf 462} (1996) 563. 

\bibitem{moore2}
  S.~Ghigna et al., Astrophys.\ J.\  {\bf 544} (2000) 616. 


\bibitem{primack}
  J.R.~Primack, astro-ph/0205391.

\bibitem{gs}
  P.~Gondolo and J~Silk, Phys. Rev. Lett. {\bf 83} (1999) 1719.

\bibitem{bhullio} 
   P.~Ullio, H.~Zhao and M.~Kamionkowski,
   Phys.\ Rev.\ D {\bf 64} (2001) 043504.

\bibitem{bu}
   L.~Bergstr{\"o}m and P.~Ullio, Nucl.\ Phys.\ {\bf B504} (1997) 27.

\bibitem{ub}
   P.~Ullio and L.~Bergstr{\"o}m, Phys.\ Rev.\ {\bf D57} (1998) 1962.
 
\bibitem{pohlradio}   
  M.~Pohl, Astron. Astrophys. {\bf 317} (1997) 441.

\bibitem{hu-nielsen} 
   H. Hu, J. Nielsen, Wisc-Ex-99-352, 1999

\bibitem{read} 
   A. L. Read, DELPHI 97-158 PHYS 737, 1997


\bibitem{msugra}
   L.~J.~Hall, J.~Lykken and S.~Weinberg,
   Phys.\ Rev.\ D {\bf 27} (1983) 2359.

\bibitem{feng}  
   J.~L.~Feng, K.~T.~Matchev and F.~Wilczek, Phys. Rev. D {\bf 63} (2001) 
   045024.

\bibitem{msd5}
   A. Bottino, F. Donato, N. Fornengo and S. Scopel, 
   Phys. Rev. {\bf D63} (2001) 125003.

\bibitem{msd4}
   J.~Ellis, J. L. Feng, A. Ferstl, K. T. Matchev and K. A. Olive, 
   Eur. Phys. J. {\bf C24} (2002) 311-322

\bibitem{msd3}
   J. Ellis, A. Ferstl and K. A. Olive,
   Phys. Lett. {\bf B532} (2002) 318.

\bibitem{msd2}
   V. Bertin, E. Nezri and J. Orloff, Eur. Phys. J. {\bf C26} (2002) 111.

\bibitem{msd1}
   U.~Chattopadhyay, A.~Corsetti and P.~Nath, hep-ph/0303201.

\bibitem{isasugra} 
   H.~Baer, F.~E.~Paige, S.~D.~Protopopescu and X.~Tata,
   hep-ph/0001086. \\
   The source code is available at ftp://ftp.phy.bnl.gov/pub/isajet

\bibitem{pdg02} 
   K. Hagiwara et al., Phys.~Rev. {\bf D66} (2002) 010001.

\bibitem{newcoann}
   J.~Edsjo, M.~Schelke, P.~Ullio and P.~Gondolo,
   JCAP {\bf 0304} (2003) 001
   [arXiv:hep-ph/0301106].

\bibitem{ellisstau} 
  J.R. Ellis, T. Falk, K.A. Olive, Phys.~Lett. {\bf B444} (1998) 367.

\bibitem{focus1}
   J.~L.~Feng and T.~Moroi, Phys.\ Rev.\ {\bf D61} (2000) 095004.

\bibitem{focus2}
   J.~L.~Feng, K.~T.~Matchev and T.~Moroi, Phys.\ Rev.\ Lett.\ 
   {\bf 84} (2000) 2322.


\end{thebibliography}
\end{document}